\documentclass[12pt]{article}

\usepackage{array}
\usepackage{epsfig}
\usepackage{amsmath}
\usepackage{amssymb}
\usepackage{graphics,graphpap}
\usepackage{cite}

\renewcommand{\thefootnote}{\fnsymbol{footnote}}

\def\s#1{\setbox0=\hbox{$#1$}%
  \rlap{\ifdim\wd0>.7em\kern.22\wd0\else\kern.1\wd0\fi /}#1}


\begin{document}

\begin{titlepage}
\begin{flushright}
\begin{tabular}{l}
IPPP/16/23\\
\end{tabular}
\end{flushright}
\vskip1.5cm
\begin{center}
{\Large \bf \boldmath
On the ultimate precision of meson mixing observables
}
\vskip1.3cm 
{\sc
Thomas Jubb                   \footnote{thomas.jubb@durham.ac.uk}$^{,1}$,
Matthew Kirk                  \footnote{m.j.kirk@durham.ac.uk}$^{,1}$,
Alexander Lenz                \footnote{alexander.lenz@durham.ac.uk}$^{,1}$,
Gilberto Tetlalmatzi-Xolocotzi\footnote{gilberto.tetlalmatzi-xolocotz@durham.ac.uk}$^{,1}$}
  \vskip0.5cm
        $^1$ IPPP, Department of Physics,
Durham University, Durham DH1 3LE, UK \\

\vskip2cm

{\em Version of \today}

\vskip3cm

{\large\bf Abstract\\[10pt]} \parbox[t]{\textwidth}{
Meson mixing is considered to be an ideal testing ground for new physics searches.
Experimental precision has greatly increased over the recent years, exceeding in
several cases the theoretical precision.
A possible limit in the theoretical accuracy could be a hypothetical
breakdown of quark-hadron duality. 
We propose a simple model for duality violations and give stringent phenomenological
bounds on its effects on mixing observables, indicating regions where  future measurements of  $\Delta \Gamma_d$, $a_{sl}^d$ and $a_{sl}^s$ would 
give clear signals of new physics.
Finally, we turn our attention to the charm sector, and reveal that a modest duality violation of about 20\%
could explain the huge difference between HQE predictions for $D$ mixing and experimental data.
}

\vfill

\end{center}
\end{titlepage}

\setcounter{footnote}{0}
\renewcommand{\thefootnote}{\arabic{footnote}}
\renewcommand{\theequation}{\arabic{section}.\arabic{equation}}

\newpage

\section{Introduction}
\label{Introduction}
\setcounter{equation}{0}

Despite having passed numerous tests, the standard model of particle physics
leaves many fundamental questions unanswered which might be resolved by extensions
of this model. Flavour physics is an ideal candidate for general indirect
new physics searches, as well as for dedicated CP-violating studies, which might shed light
on the unsolved problem of the baryon asymmetry in the Universe.
For this purpose hadronic uncertainties on flavour observables have to be under control.
Various flavour experiments have achieved a high precision in many observables, in several cases challenging
the precision of theory calculations.
LHCb in particular, as an experiment designed to study beauty and charm physics, contributes to the currently impressive status of
experimental precision.
As we attempt to test the SM to the highest level of precision, the question of how sure we can be about any deviations from the
current theoretical predictions being evidence of new physics comes to the fore.
Such a question is the subject we tackle in this paper.
\\
Many current theory predictions rely on the Heavy Quark Expansion (HQE), and we will examine how the idea of quark-hadron duality --
which is assumed by the HQE -- can be tested.
We will use current data from $B$ mixing, the dimuon asymmetry, and $B$ meson lifetimes to constrain violations of quark-hadron duality,
and then see how this affects the predicted values of other observables.
We also investigate how  the current trouble with inclusive predictions of mixing in the charm sector can be explained
through a mild violation of quark-hadron duality.
\\
We discuss  what improvements could be made in both theory and experiment in order to
further constrain duality violating effects, and what level of precision would be necessary to properly distinguish between
genuine new physics and merely a non-perturbative contribution to the SM calculation.
In this spirit, we also provide a first attempt at improving the theory prediction, using the latest results and aggressive error
estimates to see how theory uncertainties could reduce in the near future.
\\
Our paper is organised as follows: in Sec.~\ref{dualityviolation} we explain the basic ideas of duality violation
in the HQE. We introduce in Sec.~\ref{Bmixing} a simple parameterisation for duality violation in $B$ mixing
and we derive bounds on its possible size. The dimuon asymmetry and the lifetime ratio $\tau(B_s^0) / \tau (B_d^0)$ can provide complementary bounds
on duality violation, which is discussed in  Sect.~\ref{dimuon} and Sect.~\ref{lifetimeratio}.
The bounds in the $B$ system depend strongly on the theory uncertainties, hence we present in
Sect.~\ref{numericalupdate} a numerical update of the mixing observables with an aggressive error estimate for the
input parameters.
In Sec.~\ref{Dmixing} we study possible effects of duality violation in D-mixing.
We conclude in Sec.~\ref{summary} with a short summary and outlook.
The appendices contain more details to the studies in the main text.

\section{Duality violation}
\label{dualityviolation}
\setcounter{equation}{0}
In 1979 the notion of duality  was introduced by Poggio, Quinn and Weinberg \cite{Poggio:1975af} for the process
$e^+ + e^- \to $ {\it hadrons}.\footnote{The concept of duality was already used in 1970 for electron proton scattering by Bloom and Gilman
\cite{Bloom:1970xb,Bloom:1971ye}.}
 The basic assumption is that this process can be well approximated by
a quark level calculation of $e^+ + e^- \to q + \bar{q}$.
In this work we will investigate duality in the case of decays of heavy hadrons, which are described by 
the heavy quark expansion
(see e.g. \cite{Khoze:1983yp,Shifman:1984wx,Bigi:1991ir,Bigi:1992su,Blok:1992hw,Blok:1992he,Chay:1990da,Luke:1990eg} 
for pioneering papers and \cite{Lenz:2014jha} for a recent review). 
The HQE is a systematic expansion of the decay rates of $b$ hadrons in inverse powers of the heavy quark mass.
\begin{equation}
\Gamma = \Gamma_0 
+ \frac{\Lambda^2}{m_b^2} \Gamma_2
+ \frac{\Lambda^3}{m_b^3} \Gamma_3
+ \frac{\Lambda^4}{m_b^4} \Gamma_4
+ ... \; ,
\end{equation}
with $\Lambda$ being of the order of the hadronic scale.\footnote{One gets different values of $\Lambda$ for different observables.
The numerical value of $\Lambda$ has to be determined by an explicit calculation. For the case of $\Delta \Gamma_s$ one gets e.g.
$\Lambda / m_b \approx 1/5$  \cite{Lenz:2011zz} and thus $\Lambda \approx 1$ GeV.}
One finds that there are no corrections of order $\Lambda/m_b$ and that some corrections
from the order $\Lambda^3/m_b^3$ onwards are enhanced by an additional phase space factor of $16 \pi^2$.
The HQE assumes quark hadron duality, i.e. that the hadron decays can be described at the quark level.
A violation of duality could correspond to non-perturbative terms like $ \exp [ - m_b/\Lambda]$, which give 
vanishing contributions, when being Taylor expanded around $\Lambda / m_b = 0$ (see e.g. \cite{Shifman:2000jv} and also
\cite{Bigi:2001ys} for a detailed discussion of duality, its violations and some possible models for duality violations).
To estimate the possible size of these non-perturbative terms we note first that the
actual expansion parameter of the HQE is not  $\Lambda/m_b$ but the hadronic scale $\Lambda$ normalised to the momentum release 
$\sqrt{M_i^2 - M_f^2}$, where $M_i$ is the mass of the initial state and $M_f$ the sum of the final state masses. This can be shown
by an explicit derivation of the HQE.
Hence the expansion parameter for the quark-level decay $b \to c \bar{c} s$, $\Lambda / \sqrt{m_b^2- 4 m_c^2}$, is considerably 
larger than for the decay $b \to u \bar{u} u$, where it is $\Lambda / m_b$. In other words: the less phase space is accessible in the final state,
the worse is the convergence property of the HQE for this class of decays and the larger might be the hypothetical duality violating
terms. The remaining phase space for $B_s^0$ decay into light mesons (e.g. $B_s^0 \to K^- \pi^+$, via $b \to u \bar{u} d$), 
due to the dominant quark level decay (e.g. $B_s^0 \to D_s^- \pi^+ $, via $b \to c \bar{u} d$)
and into the leading contribution to $\Delta \Gamma_s$ (e.g. $B_s^0 \to  D_s^+ D_s^- $, via $b \to c \bar{c} s$) reads
\begin{eqnarray} 
M_{B_s^0} - M_K - M_\pi       & = & 4.73 \, {\rm GeV} \, , 
\\
M_{B_s^0} - M_{D_s^+} - M_\pi & = & 3.26 \, {\rm GeV} \, , 
\\
M_{B_s^0} - 2 M_{D_s^{(*)+}}       & = & 1.43 (1.15)\, {\rm GeV} \, .
\end{eqnarray}
The crucial question is, whether the phase space in $B_s^0 \to  D_s^{(*)+} D_s^{(*)-} $ is still large enough to ensure quark hadron duality.
\\
To get some idea for the possible values of the expansion parameter and the non-perturbative terms in inclusive $b$-quark decays, we 
vary $\Lambda$ within $0.2$ and $2$ GeV,\footnote{This is twice the scale one finds in $\Delta \Gamma_s$\cite{Lenz:2011zz}.} 
$m_b$ within $4.18$ and $4.78$ GeV and $m_c$ within  $0.975$ and $1.67$ GeV: 
  \begin{equation}
  \begin{array}{|c||c|c|c|}
  \hline
  \mbox{Channel}    & \mbox{Expansion parameter } x & \mbox{Numerical value} & \exp[-1/x]
  \\
  \hline
  \hline
  b \to c \bar{c} s & \frac{\Lambda}{\sqrt{m_b^2 - 4 m_c^2}} 
  \approx  \frac{\Lambda}{m_b} \left( 1 + 2 \frac{m_c^2}{m_b^2} \right) & 0.054-0.58 & 9.4 \cdot 10^{-9} - 0.18
 \\
 \hline
 b \to c \bar{u} s & \frac{\Lambda}{\sqrt{m_b^2 - m_c^2}} 
 \approx  \frac{\Lambda}{m_b} \left( 1 + \frac12 \frac{m_c^2}{m_b^2} \right) & 0.045 - 0.49& 1.9 \cdot 10^{-10} - 0.13
 \\
 \hline
 b \to u \bar{u} s & \frac{\Lambda}{\sqrt{m_b^2 - 4 m_u^2}} 
 =  \frac{\Lambda}{m_b}  & 0.042 - 0.48 & 4.2 \cdot 10^{-11} - 0.12
 \\
 \hline
 \end{array}
 \end{equation}
From this simple numerical exercise one finds that duality violating terms could easily be of a similar size as
the expansion parameter of the HQE. Moreover decay channels like $b \to c \bar{c} s$ might be more strongly affected 
by duality violations compared to e.g. $b \to u \bar{u} s$. This agrees with the naive expectation that decays with a smaller
final state phase space might be more sensitive to duality violation. 
\\
Obviously duality cannot be proved directly, because this would require a complete solution of 
QCD and a subsequent comparison with the HQE expectations, which is clearly not possible.
To make statements about duality violation in principle two strategies can be performed:
\begin{itemize}
\item[a)] Study simplified models for QCD, e.g. the t'Hooft model (a two-dimension model for QCD, see e.g. 
          \cite{Shifman:2000jv,Bigi:2001ys,Grinstein:2001zq,Grinstein:2001ep,Lebed:2000gm,Bigi:1998kc}) 
          and develop models 
          for duality violations, like instanton-based and resonance-based models 
          (see e.g. \cite{Shifman:2000jv,Bigi:2001ys}). 
\item[b)] Use a pure phenomenological approach, by comparing experiment with HQE predictions.
\end{itemize}
In this work we will follow strategy b) and use a simple parameterisation for duality violation in mixing observables
and lifetime ratios, which will be most pronounced for the $b \to c \bar{c} s$ channel.
At this stage it is interesting to note that for many years there have been
problems related to  applications of the HQE for inclusive $b$-hadron decays and most of them
seemed to be related to the $b \to c \bar{c} s$ channel:
\begin{itemize}
\item For quite some time the experimental value for the $\Lambda_b$ lifetime was considerably lower than early theory predictions 
      \cite{Shifman:1986mx}, which indicated a value quite close to the $B_d^0$ lifetime (see \cite{Lenz:2014jha} for a detailed
      review). These results triggered theoretical attempts to explain the discrepancy with a failure of the HQE, 
      see e.g. the discussion in \cite{Altarelli:1996gt}, where a simple model for a modification of the HQE was 
      suggested in order to explain experiment, see also \cite{Uraltsev:1996ta} and \cite{Neubert:1996we}.
      The dominant contribution to the  $\Lambda_b$ lifetime is given by the $b \to c \bar{u} d$ and  $b \to c \bar{c} s$
      transitions. To a large extent the $\Lambda_b$ lifetime problem  has now been solved experimentally, see the detailed 
      discussion in \cite{Lenz:2014jha}, mostly by new measurements from LHCb \cite{Aaij:2013oha,Aaij:2014zyy,Aaij:2014owa}
      and the new data confirm the early theory estimates \cite{Shifman:1986mx}.
      However, there is still a large theory uncertainty remaining due to unknown non-perturbative matrix 
      elements that could be calculated with current lattice-QCD techniques. 
\item For quite some time the values of the inclusive semi-leptonic branching ratio of $B$ mesons as well as the average
      number of charm quarks per $b$-decay (missing charm puzzle) disagreed between experiment and theory, see e.g.
      \cite{Bigi:1993fm,Falk:1994hm,Buchalla:1995kh,Lenz:2000kv}. 
      Modifications of the decay   $b \to c \bar{c} s$ were considered as a potential candidate for solving this problem.
      This issue has been improved considerably by new data and and new calculations \cite{Krinner:2013cja}.
      Again, there still a considerable uncertainty remains due to unknown NNLO-QCD corrections. First estimates
      suggest that such corrections could be large \cite{Czarnecki:2005vr}.
\item Because of a cancellation of weak annihilation contributions it is theoretically expected (based on the HQE) that the 
      $B_s^0$ lifetime is more or less equal to the $B_d^0$ lifetime, see e.g. an early estimate from 1986 \cite{Shifman:1986mx} 
      or the review \cite{Lenz:2014jha} for updated values.
      For quite some time experiment found deviations of $\tau (B_s^0) / \tau (B_d^0)$ from one - we have plotted the experimental
      averages from HFAG \cite{Amhis:2014hma} from 2003 onwards in Fig.\ \ref{fig:lifetime_history}. 
      Currently there is still a small difference between data and the HQE prediction, which  will be discussed further 
      Section~\ref{lifetimeratio}.
      Here again a modification of the  $b \to c \bar{u} d$ and/or the  $b \to c \bar{c} s$ transitions might solve the problem.
      \begin{figure}[htp]
      \includegraphics[width=0.9\textwidth]{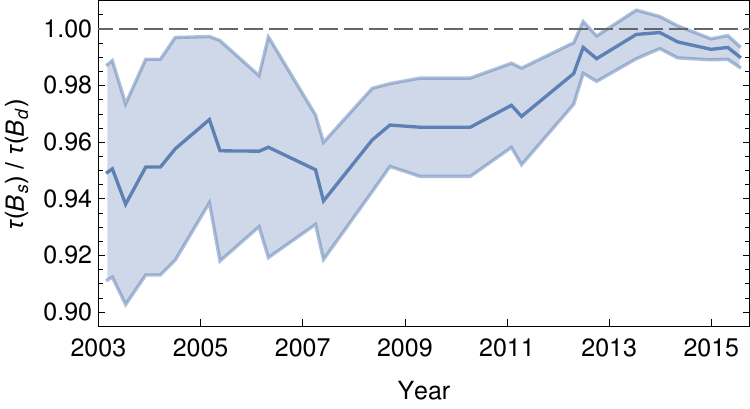}
         \caption{Historical values of the lifetime ratio \(\tau(B_s^0) / \tau(B_d^0)\) as reported by HFAG \cite{Amhis:2014hma} since 2003.
         The solid line shows the central value and the shaded line indicates the \(1 \sigma\) region, the dotted line corresponds to the 
         theory prediction, which is essentially one, with a tiny uncertainty.
         }
\label{fig:lifetime_history}
\end{figure}
\item The large observed value of the dimuon asymmetry \cite{Abazov:2010hv,Abazov:2010hj,Abazov:2011yk,Abazov:2013uma}
      could not only have hinted towards  new physics but also to large values of $\Gamma_{12}^s$, which is dominated by $b \to c \bar{c} s$.
      Thus it was suggested to investigate the dimuon asymmetry without making use of the theory prediction of $\Gamma_{12}^s$ 
      \cite{Ligeti:2010ia}, which was criticised in \cite{Lenz:2011zz}.
      The issue of the dimuon asymmetry is still not settled and we will discuss it further below.
\end{itemize}
All of these problems are currently considerably softened and huge duality violations are now ruled out by experiment \cite{Lenz:2012mb},
in particular by the measurement of  the decay rate difference of neutral $B_s^0$ mesons, $\Delta \Gamma_s$, 
which is to a good approximation a $b \to c \bar{c} s$ transition.
But there is still space for a small amount of duality violation - which will be quantified in this work.
\\
We will thus investigate the decay rate difference  $\Delta \Gamma_s = 2 |\Gamma_{12}^s| \cos \phi_{12}^s$ (see e.g. \cite{Artuso:2015swg} for the
basic mixing formulae) in more detail.
The off diagonal matrix element, $\Gamma_{12}^s$, of the $B_s^0$ meson mixing matrix can be determined as 
a double insertion of the effective weak Hamiltonian describing weak decays of $B_s^0$ mesons:
\begin{eqnarray}
\Gamma_{12}^s & = &  \frac{1}{M_{B_s^0}} \Im 
\left[ 
\frac{i}{2} \int d^4x \langle B_s^0 | T \left\{ {\cal H} (x)  {\cal H} (0) \right\}
                                | \bar{B}_s^0\rangle
\right] \, .
\label{Gamma12}
\end{eqnarray}
According to the HQE  this expression can be expanded in powers of $\Lambda / m_b$
\begin{equation}
\Gamma_{12}^s = \frac{\Lambda^3}{m_b^3}
\left(\Gamma_3^{s,(0)} + \frac{\alpha_s}{4 \pi} \Gamma_3^{s,(1)} + ... \right)
+
\frac{\Lambda^4}{m_b^4}
\left(\Gamma_4^{s,(0)} + ... \right) + ... \, .
\label{HQE}
\end{equation}
The leading term $\Gamma_3^{s,(0)}$ has been calculated quite some time ago by
\cite{Ellis:1977uk,Hagelin:1981zk,Franco:1981ea,Chau:1982da,Buras:1984pq,Khoze:1986fa},
NLO-QCD corrections  $\Gamma_3^{s,(1)}$ have been determined in  \cite{Beneke:1998sy,Ciuchini:2003ww,Beneke:2003az}
and  sub-leading mass corrections were done in \cite{Beneke:1996gn,Dighe:2001gc,Badin:2007bv}.
Corresponding lattice values were determined by \cite{Becirevic:2001xt,Bouchard:2011xj,Carrasco:2013zta,Dowdall:2014qka}.
\\
The most recent numerical update for the mixing quantities is given in \cite{Artuso:2015swg} 
(superseding the numerical predictions in \cite{Lenz:2006hd,Lenz:2011ti})
and can be compared to the
experimental values from e.g. HFAG \cite{Amhis:2014hma}. The theory prediction uses conservative ranges for the input parameters - 
we will present a more aggressive estimate in Sec.~\ref{numericalupdate} and in Appendix \ref{app:D}.
\begin{equation}
\begin{array}{|c||c|c|}
\hline
\mbox{Observable} &  \mbox{SM}                            & \mbox{Experiment}
\\
\hline
\hline
\Delta M_s        & ( 18.3 \pm 2.7)\,  \mbox{ps}^{-1 }   & ( 17.757 \pm 0.021) \, \mbox{ps}^{-1}
\\
\hline
\Delta \Gamma_s   & ( 0.088 \pm 0.020) \, \mbox{ps}^{-1 }& ( 0.082 \pm 0.006) \, \mbox{ps}^{-1}
\\
\hline
a_{sl}^s          & ( 2.22 \pm 0.27) \cdot 10^{-5}       & ( 170 \pm 300) \cdot 10^{-5}
\\
\hline
\Delta \Gamma_s /
\Delta M_s        & 48.1 \, (1  \pm 0.173) \cdot 10^{-4} & 46.2 \, (1 \pm 0.073) \cdot 10^{-4}
\\
\hline
\hline
\Delta M_d        & ( 0.528 \pm 0.078)\,  \mbox{ps}^{-1 }   & ( 0.5055 \pm 0.0020) \, \mbox{ps}^{-1}
\\
\hline
\Delta \Gamma_d   & ( 2.61 \pm 0.59) \cdot 10^{-3} \, \mbox{ps}^{-1 }& 0.66 ( 1\pm 10) \cdot 10^{-3} \, \mbox{ps}^{-1}
\\
\hline
a_{sl}^d          & (-4.7 \pm 0.6) \cdot 10^{-4}       & ( -15 \pm 17) \cdot 10^{-4}
\\
\hline
\Delta \Gamma_d /
\Delta M_d        & 49.4 \, (1  \pm 0.172) \cdot 10^{-4} & 13  \, (1 \pm 10) \cdot 10^{-4}
\\
\hline
\end{array}
\label{comparison1}
\end{equation}
The experimental average for $a_{sl}^s$ has been taken from \cite{Aaij:2016yze}.
Experiment and theory agree very well for the quantities $\Delta M_q$ and $\Delta \Gamma_s$. The semileptonic asymmetries and the
decay rate difference in the $B_d^0$ system have not been observed yet.
More profound statements about the validity of the theory  can be made by comparing the ratio
of $\Delta \Gamma_s$ and $\Delta M_s$, where some theoretical uncertainties cancel and we get 
\begin{equation}
\frac{\left( \frac{\Delta \Gamma_s}{\Delta M_s} \right)^{\rm Exp}}{\left( \frac{\Delta \Gamma_s}{\Delta M_s}\right)^{\rm SM}} 
= 0.96 \pm 0.22 \,  \, \, ( \mbox{at} \, \, 95 \% \, \, \mbox{C.L.}).
\label{testHQE}
\end{equation}
The central value shows a very good agreement of experiment and HQE predictions. The remaining uncertainty leaves
some space for new physics effects or for violations of duality. We have taken here the 2 $\sigma$ range of  the experimental value,
while we consider the theory range to cover all allowed values.
Thus  we conclude that 
in the most sensitive decay channel, $b \to c \bar{c} s$, duality seems to be violated by at most $22\%$. In the next chapter we try to
investigate these possibilities a little more in detail.
\\
Assuming no new physics in $\Delta M_s$ we can further reduce the theory error for $\Delta \Gamma_s$ by
using\footnote{In the ratio $\Delta \Gamma_s / \Delta M_s$ theoretical uncertainties are cancelling and thus the corresponding
theory error is smaller than for $\Delta \Gamma_s$ alone. We would reintroduce this uncertainty by multiplying with
the theory value of $\Delta M_s$. Using instead the experimental value of $\Delta M_s$, which has in comparison a negligible
error we get a more precise prediction of $\Delta \Gamma_s$, which, however, only holds under the assumption that $\Delta M_s^{\rm Exp}$ is
given by its standard model value. Therefore we denote this quantity with an superscript ``SMb''.}
\begin{eqnarray}
\Delta \Gamma_s^{2015, \rm SMb} = 
\left(  \frac{\Delta \Gamma_
s}{\Delta M_s} \right)^{\rm SM} \cdot \Delta M_s^{\rm Exp}  = 
0.085 \pm 0.015  \, \mbox{ps}^{-1} \, .
\end{eqnarray}
This is currently the most precise prediction for the decay rate difference; in Section~\ref{numericalupdate} 
we will give a less conservative 
estimate of the SM prediction for $\Delta \Gamma_s$, with an even smaller uncertainty.
\\
Alternatively to Eqs.(\ref{Gamma12}),(\ref{HQE}), $\Delta \Gamma_s$ can in principle be determined from exclusive decays, 
avoiding thus the expansion in $\Lambda/m_b$:
\begin{eqnarray}
\Delta \Gamma_s & = & \Gamma_{B_L} - \Gamma_{B_H}
= \sum \limits_f \left| \langle f|{\cal H} | B_L \rangle  \right|^2 
                    - \sum \limits_f \left| \langle f|{\cal H} | B_H \rangle  \right|^2
= 4 {\Re} \left[ pq^* \sum \limits_f  \langle \bar{B}_s^0 | {\cal H} \ f \rangle \langle f | {\cal  H} | B_s^0   \right] \, ,
\nonumber
\\
\label{exclusive}
\end{eqnarray}
where $f$ denotes final states common to $B_s^0$ and $\bar{B}_s^0$. The coefficients $p$ and $q$ describe the basis change from 
$B_s^0$, $\bar{B}_s^0$ to $B_L$, $B_H$ (see e.g. \cite{Artuso:2015swg} for the
basic mixing formulae). However, the theoretical determination of decay rates for exclusive $B_s^0$ decays is in general 
an unsolved hadronic problem, but for certain cases (not including e.g. $B_s \to D_s^{+(*)}  D_s^{-(*)}$!)
factorisation seems to be applicable, see e.g.
\cite{Beneke:1999br,Beneke:2000ry,Beneke:2001ev}.
Using Eq.(\ref{exclusive}), $\Delta \Gamma_s$ can be estimated by summing up exclusive branching ratios, assuming naive factorisation. 
This approach has been followed  e.g. in  \cite{Aleksan:1993qp} and \cite{Chua:2011er}. Using LO-QCD expressions and taking into account 
a certain number of 2- and 3-body decays the authors of \cite{Chua:2011er} obtained
\begin{equation}
\Delta \Gamma_s^{\rm exclusive} = (0.111 \pm 0.057) \, {\rm ps}^{-1} \, ,
\end{equation}
which is consistent with the direct measurement and with the HQE determination, but suffers from much larger uncertainties. 
\\
From a theoretical point of view there is an interesting limiting case, which should, however, not be used for phenomenological applications:
$m_c \to \infty$, $m_b \to 2 m_c$ (Shifman-Voloshin limit \cite{Shifman:1987rj}) and neglecting certain terms of order $1/N_c$ one gets
\cite{Aleksan:1993qp,Dunietz:2000cr}:
\begin{equation}
2 Br (D_s^{(*)+}D_s^{(*)-})^{\rm SV-limit} = \Delta \Gamma_s^{\rm SV-limit} \frac{\tau (B_s^0)}{ \cos(\phi_{12}^s)}
\, ,
\label{wrong} 
\end{equation}
which is, however, not full-filled to a high precision by experiment. Using the experimental result for the branching ratios 
\cite{Olive:2016xmw} 
as an input in  Eq.(\ref{wrong}) we get a decay rate difference of (see also \cite{Aleksan:1993qp,Dunietz:2000cr})
\begin{equation}
\Delta \Gamma_s^{\rm SV-limit} \leq (0.060 \pm 0.019) \, {\rm ps}^{-1} \, ,
\end{equation}
which is considerably lower than the direct determination.
On the other hand this result shows that $D_s^{(*)+}D_s^{(*)-}$ final states give the dominant contribution to $\Delta \Gamma_s$.
\\
Using the above limit and setting further $\alpha_s =0$, Aleksan et al.  \cite{Aleksan:1993qp} could show that
both the HQE approach and the exclusive approach yield analytically the following result
\begin{eqnarray}
\Delta \Gamma_s^{\rm Aleksan} & = & 
\frac{G_F^2 m_b^2 M_{B_s^0} f_{B_s}^2}{4 \pi} \left| V_{cb}^* V_{cs} \right|^2 \sqrt{1 - 4 \frac{m_c^2}{m_b^2}}
\approx  0.13 \, {\rm ps}^{-1} \, .
\label{duality}
\end{eqnarray}
This value is now considerably above the direct measurement.
Despite looking like a proof of duality, we would like to add some critical comments:
we can reproduce Eq.(\ref{duality}) from the full HQE expressions for $\Delta \Gamma_s$ by taking only the leading CKM structure
into account, by setting $\alpha_s$ to zero, by setting the bag parameters to one (for the $S-P$ operator we also set $M_{B_s} = m_b + m_s$)
and we neglect some terms of order $m_c^2/m_b^2$, while keeping it in the square root in Eq.(\ref{duality}).
These approximations lead to the effect that the result of  Eq.(\ref{duality}) is more than $50\%$ larger than the full theory prediction. 
This deviation is much larger than the current experimental and theoretical uncertainties in $\Delta \Gamma_s$. Thus we conclude that
the result of \cite{Aleksan:1993qp} has no practical relevance for our aim of constraining possible sizes of duality violation.

\subsection{$B$ mixing}
\label{Bmixing}

\begin{figure}[htp]
\includegraphics[width=0.8\textwidth]{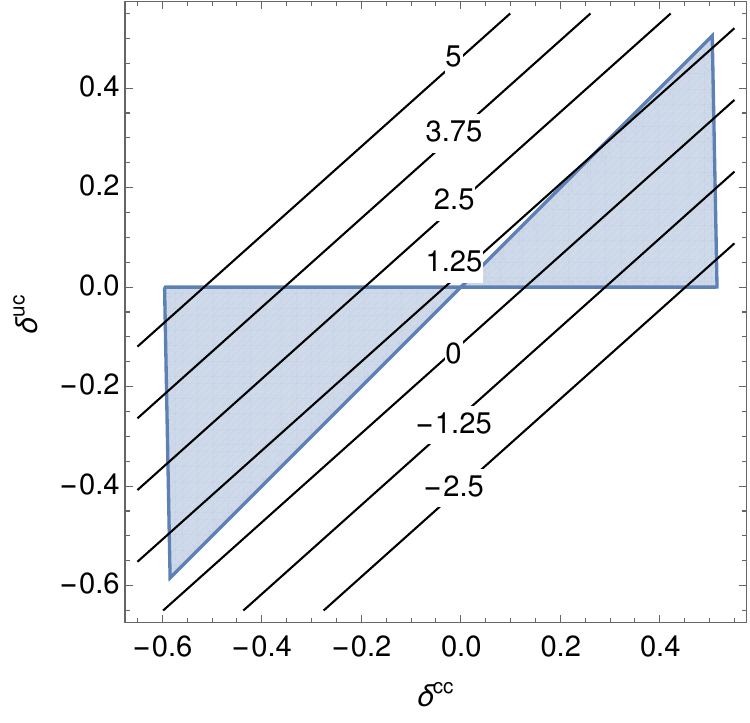}
\caption{95\% confidence limits on $\delta^{cc}$ and $\delta^{uc}$ for the $B_s^0$ system from a comparison of the experimentally 
         allowed region of $(\Delta \Gamma_s / \Delta M_s)$ with the theory expression in Eq.(\ref{DGDMs}). 
         The allowed regions for the $\delta$s are shaded blue(grey). A deviation of the 
         $\delta$s from zero will also affect the theory prediction of $a_{sl}^s$ in Eq. \ref{asls}. 
         The modification factors of $a_{sl}^s/a_{sl}^{s, \rm SM}$ are denoted by the black lines.
         }
\label{fig:bs_general_limits}
\end{figure}

\begin{figure}[htp]
      \includegraphics[width=0.44\textwidth]{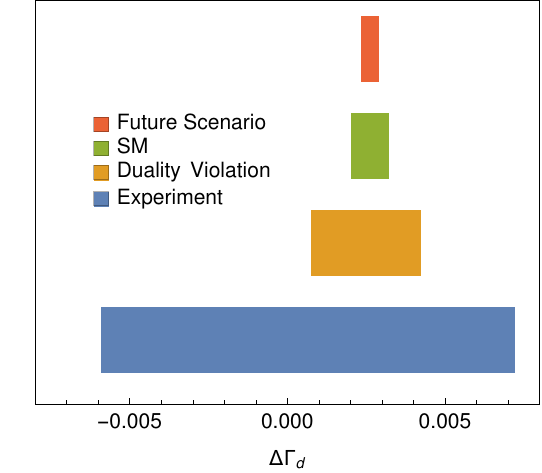}
      \hfill
      \includegraphics[width=0.52\textwidth]{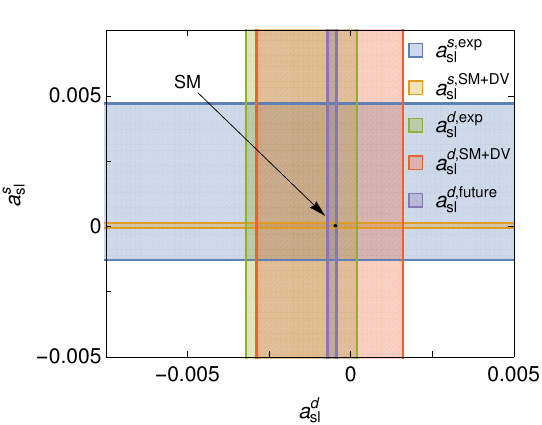}

         \caption{Comparison of SM prediction (green), SM + duality violation (brown), 
                   SM + duality violation in future (orange) and current experimental (blue)
                  bound for $\Delta \Gamma_d$ (l.h.s.). On the r.h.s. the experimental bounds on $a_{sl}^d$ (green) and $a_{sl}^s$ (blue) 
                  are shown in comparison to their theory values. The uncertainties of the SM predictions are too small to be resolved,
                  the regions allowed by duality violation are shown in brown  ($a_{sl}^s$) and orange ($a_{sl}^d$).
                  Any measurement outside these duality  allowed theory regions will
                  be a clear indication for new physics. 
                  For $a_{sl}^d$ the duality allowed region (orange) has a pronounced overlap with the experimental one, 
                  in future this region could be shrinking
                  to the dark blue region.
                  The theory uncertainties for the future duality region of $a_{sl}^s$  are so small, that they cannot be resolved in the plot. 
                  }
\label{fig:duality_bounds}
\end{figure}

The off-diagonal elements $\Gamma_{12}^s$ and $M_{12}^s$ of the mixing matrix for $B_s^0$ mesons can be expressed as
\begin{eqnarray}
\Gamma_{12}^s =  - \sum \limits_{x=u,c} \sum \limits_{y=u,c} \lambda_x \lambda_y \Gamma_{12}^{s,xy} \; ,
&&
M_{12}^s      =  \lambda_t^2 \tilde{M}_{12}^s \; .
\label{HQE2}
\end{eqnarray}
Here we have separated the CKM dependence, $\lambda_q = V_{qs}^* V_{qb}$.  
$ \Gamma_{12}^{s,xy}$ describes the on-shell part of a $B_s^0$ mixing diagram with internal
$x$ and $y$ quarks, $x,y \in \{u, c\}$, and  $\tilde{M}_{12}^s$ describes the off-shell part without CKM factors.
\\
For simplicity we give only the expressions for $B_s^0$ mesons when modifications for $B_d^0$ mesons 
are obvious, and we will explicitly present expressions for the $B_d^0$ sector only when  they are 
non-trivial.
The physical observables $\Delta M_s$, $\Delta \Gamma_s$ and $a_{sl}^s$ are related to the ratio
$\Gamma_{12}^s/M_{12}^s$ - there several theory uncertainties are cancelling - via
\begin{eqnarray}
\frac{\Delta \Gamma_s}{\Delta M_s} & = & - {\Re} \left(\frac{\Gamma_{12}^s}{M_{12}^s} \right) \; ,
\hspace{1cm}
a_{sl}^s                            =    {\Im} \left(\frac{\Gamma_{12}^s}{M_{12}^s} \right) \; .
\end{eqnarray}
Using the unitarity of the CKM matrix we can simplify $\Gamma_{12}^s / M_{12}^s$.
\begin{eqnarray}
      - \frac{\Gamma_{12}^s}{M_{12}^s}  & = & 
       \frac{\Gamma_{12}^{s,cc}}{\tilde{M}_{12}^s}
        + 2 \frac{\lambda_u}{\lambda_t}
          \frac{\Gamma_{12}^{s,cc} -\Gamma_{12}^{s,uc}}{\tilde{M}_{12}^s}
      \label{ratio1}
        + \left(\frac{\lambda_u}{\lambda_t}\right)^2
          \frac{\Gamma_{12}^{s,cc} -2 \Gamma_{12}^{s,uc}+\Gamma_{12}^{s,uu} }{\tilde{M}_{12}^s}
      \\
      & = & - 10^{-4}
       \left[ c + a \frac{\lambda_u}{\lambda_t} + b\left(\frac{\lambda_u}{\lambda_t}\right)^2
      \right] \; .
      \label{ratio2}
      \end{eqnarray}
 Eq.(\ref{ratio2}) introduces the $a$, $b$ and $c$ notation of \cite{Beneke:2003az}.
 The way of writing $\Gamma_{12}^s/M_{12}^s$ in Eq.(\ref{ratio1}) and Eq.(\ref{ratio2}) can be viewed
    as a Taylor expansion in the small CKM parameter $\lambda_u/\lambda_t$, for which we get 
    (we use the same the CKM input as \cite{Artuso:2015swg}; the values were taken in 2015 from
  CKMfitter \cite{Charles:2004jd}, similar values can be obtained from UTfit \cite{Ciuchini:2000de}.)
\begin{equation}
\begin{array}{|c||c|c|}
\hline
\mbox{CKM}                  & B_s^0                                           & B_d^0
\\
\hline
\hline
\frac{\lambda_u}{\lambda_t}                 & - 8.05 \cdot 10^{-3} + 1.81 \cdot 10^{-2} I & 7.55 \cdot 10^{-3} - 4.05 \cdot 10^{-1} I
\\
\hline
\left( \frac{\lambda_u}{\lambda_t} \right)^2& - 2.63 \cdot 10^{-4} - 2.91 \cdot 10^{-4} I &-1.64 \cdot 10^{-1} - 6.11 \cdot 10^{-3} I
\\
\hline
\end{array}
\label{CKM}
\end{equation}
    In addition to the CKM suppression  a pronounced GIM-cancellation  \cite{Glashow:1970gm} is arising
    in the coefficients $a$ and $b$ in Eq.(\ref{ratio2}). With the input parameters described
    in \cite{Artuso:2015swg} we get for the numerical values of $a$, $b$ and $c$:
\begin{equation}
\begin{array}{|c||c|c|}
\hline
           & B_s^0              & B_d^0
\\
\hline
\hline
         c &  -48.0 \pm 8.3    &  -49.5 \pm 8.5
 \\
\hline
         a &  + 12.3 \pm 1.4   & +11.7 \pm 1.3
\\
\hline
         b &  + 0.79 \pm 0.12  &+0.24 \pm 0.06
\\
\hline
\end{array}
\label{abcSM}
\end{equation}
From this hierarchy we see, that $\Delta \Gamma_q / \Delta M_q$ is given to a very good approximation by
$-0.0001 c$ and $a_{sl}^q$ by $0.0001 a \Im (\lambda_u / \lambda_t)$.
\\
Next we introduce a simple model for duality violation. Such effects are typically expected to be 
larger, if the phase-space of a $B_s^0$ decay becomes smaller. Thus $b$ quark decays into two charm quarks
are expected to be more strongly affected by duality violating effects compared to $b$ quark decays into two
up quarks. Motivated by the observations in Section \ref{dualityviolation} we write to a first approximation\footnote{Similar models
have been used in \cite{Palmer:1993kx,Dunietz:1996zn,Lenz:1997aa} for penguin insertions with a $c \bar{c}$-loop.}: 
\begin{eqnarray}
\Gamma_{12}^{s,cc} & \to & \Gamma_{12}^{s,cc} (1 + 4 \delta) \; ,
\label{adhoc1}
\\
\Gamma_{12}^{s,uc} & \to & \Gamma_{12}^{s,uc} (1 +   \delta) \; ,
\label{adhoc2}
\\
\Gamma_{12}^{s,uu} & \to & \Gamma_{12}^{s,uu} (1 + 0 \delta) \; .
\label{adhoc3}
\end{eqnarray}
The $c \bar{c}$ contribution is affected by a correction of $4 \delta$, $c \bar{u}$ by $\delta$
and $u \bar{u}$ is not affected at all. Already at this stage ones sees that such a model is softening  
GIM cancellations in the ratio $\Gamma_{12}^s / M_{12}^{s}$; we get 
\begin{equation}
\frac{\Gamma_{12}^s}{ M_{12}^s} = 10^{-4}
\left[
c ( 1+ 4 \delta) +
\frac{\lambda_u}{\lambda_t} \left( a + \delta (6c + a) \right) +
 \frac{\lambda_u^2}{\lambda_t^2} (b + \delta (2c+a))
\right] \; .
\end{equation}
Studying this expression, we find that the decay rate difference is mostly given by the first term on the r.h.s., 
so we expect $\Delta \Gamma_s / \Delta M_s \approx - c (1+4 \delta) \cdot 10^{-4}$, which is equivalent to our naive
starting point of comparing experiment and theory prediction for $\Delta \Gamma_s$. The semi-leptonic CP asymmetries
will be dominantly given by the second term on the r.h.s., $a_{sl}^s \approx {\Im} (\lambda_u / \lambda_t) 
\left[ a + \delta (6c + a) \right] \cdot 10^{-4}$. Now the duality violating coefficient $\delta$ is GIM enhanced by $(6c+a)$
compared to the leading term $a$. Having an agreement of experiment and theory for semileptonic CP asymmetries
could thus provide very strong constraints on duality violation.
Using the values of $a$, $b$ and $c$ from Eq.(\ref{abcSM}) and the CKM elements from Eq.(\ref{CKM}) we 
get for the observables $\Delta M_q$, $\Delta \Gamma_q$ and $a_{sl}^q$ the following dependence
on the duality violating parameter $\delta$:
\begin{equation}
\begin{array}{|c||c|c|}
\hline
\mbox{Observable}           & B_s^0              & B_d^0
\\
\hline
\hline
\frac{\Delta \Gamma_q}{\Delta M_q} & 48.1 (1 + 3.95 \delta) \cdot 10^{-4} 
                                   & 49.5 (1 + 3.76 \delta) \cdot 10^{-4}
 \\
\hline
\Delta \Gamma_q                    & 0.0880 (1 + 3.95 \delta)  \, \mbox{ps}^{-1}
                                   & 2.61  (1 + 3.76 \delta) \cdot 10^{-3} \, \mbox{ps}^{-1}
 \\
\hline
a_{sl}^q                         &  2.225 (1 - 22.3 \delta)  \cdot 10^{-5} &- 4.74 (1 - 24.5 \delta)  \cdot 10^{-4} 
\\
\hline
\end{array}
\label{model1}
\end{equation}
As expected we find that the duality violating parameter $\delta$ has a
decent leverage on $\Delta \Gamma_q$ and a sizeable one on $a_{sl}^q$. The expressions for $\Delta \Gamma_q$ were 
obtained by simply
multiplying the theory ratio $\Delta \Gamma_q / \Delta M_q$ with the theoretical values of the mass difference, 
as given in Eq.(\ref{comparison1}).
\\
Comparing experiment and theory for the ratio of the decay rate difference $\Delta \Gamma_s$ and 
the mass difference $\Delta M_s$ we found (see Eq.(\ref{testHQE}))
an agreement with a deviation of at most $22\%$. Thus the duality violation - i.e. the factor $1 + 3.95 \delta$ in 
Table \ref{model1} -  has to be smaller than this uncertainty: 
\begin{equation}
1 + 3.95 \delta \leq 0.96 \pm 0.22 \Rightarrow \delta \in  [-0.066, + 0.046] \: .
\label{bound1}
\end{equation}
Equivalently this bound tells us that the duality violation in the cc-channel is at most $+18.2\%$ or  $-26.3\%$, 
if the effect turns out to be negative.
If there would also be an 22\% agreement of experiment and theory for the semileptonic asymmetry $a_{sl}^s$, 
then we could shrink the bound to $\delta$ down  to 0.01. Unfortunately experiment is still far away from
the standard model prediction, see Eq.(\ref{comparison1}). However, we can turn around the argument: 
even in the most pessimistic scenario - 
i.e. having a duality violation that lifts GIM suppression - the theory prediction of $a_{sl}^s$ can be 
enhanced/diminished  at most to
\begin{eqnarray}
a_{sl}^s        & = & [-0.06,5.50] \cdot 10^{-5} \; .
\label{asls1}
\end{eqnarray}
In the $B_d^0$ system a comparison of experiment and theory for the ratio of decay rate difference and mass difference 
turns out to be tricky, since $\Delta \Gamma_d$ is not yet measured, see Eq.(\ref{comparison1}).
 If we would use the current experimental bound 
on the decay rate difference $\Delta \Gamma_d$, we would get artificially large bounds on $\delta$.
Looking at the structure of the loop contributions necessary to calculate $\Gamma_{12}^d$ and $\Gamma_{12}^s$,
one finds very similar $c \bar{c}$-,  $u\bar{c}$-, $c \bar{u}$-  and  $u \bar{u}$-contributions. 
Our duality violation model is based on the phase space differences 
of decays like $B_s^0 \to D_s D_s$ $(c\bar{c})$, $B_s^0 \to D_s K$ $(u\bar{c}), (c\bar{u})$ and $B_s^0 \to \pi K $ $(u\bar{u})$, which are very pronounced.
On the other hand we find that the phase space differences of $B_s^0$ and $B_d^0$ decays are not very pronounced, i.e.
the difference between
  e.g. $B_s^0 \to D_s D_s$ vs. $B_d^0 \to D_s D$ 
is small - compared to the above differences due to different internal quarks. Hence we  conclude that the duality violation bounds from
the $B_s^0$ system can also be applied to a good approximation to the $B_d^0$ system.
With the $B_s^0$ bound we get that the theory prediction of $a_{sl}^d$ and $\Delta \Gamma_d$ can be 
enhanced/diminished  due to duality violations at most to
\begin{eqnarray}
a_{sl}^d        & \in & [-12.4,-0.6] \cdot 10^{-4} \; ,
\label{asld1}
\\
\Delta \Gamma_d & \in & [1.96, 3.06] \cdot 10^{-3} \; \mbox{ps}^{-1} \; .
\label{DG1}
\end{eqnarray}
These  numbers can be compared to the SM values obtained in \cite{Artuso:2015swg}, see Eq.(\ref{comparison1}).
In principle any measurement of these observables outside the ranges in
Eq.(\ref{asls1}), Eq.(\ref{asld1}) and Eq.(\ref{DG1}) would be a clear indication of new physics.
New physics in $\Delta \Gamma_d $ could have the very interesting effect of reducing 
\cite{Borissov:2013wwa} the still existing discrepancy of the dimuon asymmetry measured at D0
\cite{Abazov:2010hv,Abazov:2010hj,Abazov:2011yk,Abazov:2013uma}. Currently a sizeable enhancement of $\Delta \Gamma_d$ is not 
excluded by theoretical or experimental bounds \cite{Bobeth:2014rda}. Thus it is clearly important to distinguish hypothetical
duality violating effects in $\Delta \Gamma_d$ from new physics effects.
\\
Since our conclusions (new physics or unknown hadronic effects) are quite far-reaching, we try to be as conservative as possible
and we will firstly use a more profound statistical method, a likelihood ratio test
\footnote{
For our likelihood function we use a Gaussian function
\[
\mathcal{L} = \exp \left( - \frac{(\text{theory} - \text{experiment})^2}{2 \sigma^2} \right)
\]
where, to take into account the uncertainty on both theory and experiment, we take for our error the quadrature sum, i.e.\
\[
\sigma^2 = \sigma_\text{exp}^2 + \sigma_\text{theory}^2
\]
The test we apply is \(-2 \ln \mathcal{L} / \widehat{\mathcal{L}} \leq 3.84\), where our choice of 3.84 gives a 95\% confidence 
limit on our parameters and in principle we normalise by the maximum likelihood \(\widehat{\mathcal{L}}\).
However, in our model the maximum likelihood of \(\widehat{\mathcal{L}} = 1\) is always achievable, 
and so our test reduces to simply \(-2 \ln \mathcal{L} \leq 3.84\).
}.
Our more conservative bound for $\delta$ is now supposed to be given by
\begin{equation}
-0.12 \leq \delta \leq 0.10 \, ,
\label{deltainterval}
\end{equation}
with a $95\%$ confidence level (2 standard deviations).
This more conservative statistical method doubles the allowed region for $\delta$.
Inserting these values into the predictions for $a_{sl}^{d,s}$ and $\Delta \Gamma_d$ we see 
that duality violation can give at most the following ranges for the mixing observables:
\begin{eqnarray}
a_{sl}^s        & \in & [-2.8, 8.2] \cdot 10^{-5} \, ,
\\
a_{sl}^d        & \in & [-18.7,6.9] \cdot 10^{-4} \, ,
\\
\Delta \Gamma_d & \in & [1.4,3.6] \cdot 10^{-3} \; \mbox{ps}^{-1} \; .
\end{eqnarray}
The second modification to ensure that our estimates are conservative concerns our ad-hoc 
ansatz in Eqs.(\ref{adhoc1}), (\ref{adhoc2}), (\ref{adhoc3}), where we assumed that the $cc$-part is affected by
duality violations four times as much as the $cu$-part and the $uu$-part is not affected at all; 
we can obtain more general results  with the following modification
\begin{eqnarray}
\Gamma_{12}^{s,cc} & \to & \Gamma_{12}^{s,cc} (1 + \delta^{cc}) \; ,
\\
\Gamma_{12}^{s,uc} & \to & \Gamma_{12}^{s,uc} (1 + \delta^{uc}) \; ,
\\ 
\Gamma_{12}^{s,uu} & \to & \Gamma_{12}^{s,uu} (1 + \delta^{uu}) \; ,
\end{eqnarray}
with $\delta^{cc} \geq \delta^{uc} \geq \delta^{uu}$ and the requirement that all
$\delta$s must have the same sign. Now we get for the observables
\begin{eqnarray}
\frac{\Delta \Gamma_s}{\Delta M_s} & = & 
48.1 (1 + 0.982 \delta^{cc} + 0.0187 \delta^{uc} - 0.000326 \delta^{uu}) \cdot 10^{-4} \, ,
\label{DGDMs}
\\
\Delta \Gamma_d & = & 26.1 (1 + 0.852 \delta^{cc} + 0.350 \delta^{uc} - 0.202 \delta^{uu}) \cdot 10^{-4} {\mbox{ps}^{-1}}\, ,
\\
a_{sl}^s & = & 
2.225 (1 - 7.75 \delta^{cc} + 8.67 \delta^{uc} + 0.0780 \delta^{uu}) \cdot 10^{-5} \, ,
\\
\label{asls}
a_{sl}^d & = & 
-4.74 (1 - 8.52 \delta^{cc} + 9.60 \delta^{uc} - 0.0787 \delta^{uu}) \cdot 10^{-4}  \, .
\end{eqnarray}
In the case of $\Delta \Gamma_s$, which will be used to determine the size of the duality violating $\delta$s, the 
coefficients of the $uu$ component are suppressed by more than three orders of magnitude compared to the rest
and therefore neglected. For the semileptonic CP asymmetries the $uu$ duality violating component is about two 
orders of magnitude lower than the rest, thus we neglect the $uu$ component in the following. This might lead
to an uncertainty of about  $20\%$ in the duality bounds for $\Delta \Gamma_d$, which we will keep in mind.
\\
Considering only $\delta^{cc}$ and $\delta^{uc}$ we get with the  likelihood ratio test  the 
bounds depicted in Fig.\ \ref{fig:bs_general_limits} at a $95\%$ confidence level.
Fig.\ \ref{fig:bs_general_limits} shows that a  duality violation of no more than 60\% is allowed in either $\Gamma_{cc}^s$ or 
in $\Gamma_{uc}^s$. We also see that it is in principle possible to see duality violation in $\Delta \Gamma_s$ 
but not in $a_{sl}^s$ and vice versa.
Moreover we find from the functional form of $a_{sl}^s$, that this quantity achieves a maximum (minimum) when $\delta^{uc} = 0$ and 
$ \delta^{cc} < 0 $ or ($ > 0$).
Our generalised parameterisation of duality violation gives now the most conservative bounds on the
mixing observables
\begin{eqnarray}
a_{sl}^s         & \in & [-6.7,12.5] \cdot 10^{-5} \, ,
\label{asls3}
\\
a_{sl}^d         & \in & [-29,16] \cdot 10^{-4} \, ,
\label{asld3}
\\
\Delta \Gamma_d  & \in & [0.7, 4.2] \cdot 10^{-3} \, \mbox{ps}^{-1} \, .
\label{DGd3}
\end{eqnarray}
The duality bound on $a_{sl}^d$ overlaps largely with the current experimental bound on this observable,
here a future improvement in the measurement of  $a_{sl}^d$ will give an additional bound on duality violation.
\\
We are now in a position to make a strong statement: any measurement outside this range, 
cannot be due to duality violation and it will be a strong signal for new physics.
\\
Since the ranges in Eq.(\ref{asls3}), Eq.(\ref{asld3}) and Eq.(\ref{DGd3}) are considerably 
larger than the uncertainties of the corresponding standard model prediction given 
in Eq.(\ref{comparison1}) the question of how to further shrink the 
duality bounds is arising. Currently the bound on the duality violating parameters $\delta$ come entirely from $\Delta \Gamma_s$,
 where the current experimental and theoretical uncertainty adds up to $\pm 22\%$.
Any improvement on this uncertainty will shrink the allowed regions on $\delta$. In Section 3 we will discuss
a more aggressive estimate of the theory predictions for the mixing observable, indicating that a theory uncertainty
of about $\pm 10\%$ or even  $\pm 5\%$ in $\Delta \Gamma_s / \Delta M_s$ might come into sight. Including also possible improvements
in experiment this would  indicate a region for $\delta$ that is considerably smaller than the ones given in 
Eq.(\ref{asls3}), Eq.(\ref{asld3}) and Eq.(\ref{DGd3}).
The current (and a possible future) situation is summarised in Fig.\ \ref{fig:duality_bounds}.
On the l.h.s. of Fig.\ \ref{fig:duality_bounds} $\Delta \Gamma_d$ is investigated. The current experimental bound is given by
the blue region, which can be compared to the standard model prediction (green). As we have seen above, because of still sizeable 
uncertainties in $\Delta \Gamma_s$ duality violation of up to $60\%$ can currently not be excluded - this would lead to an 
extended region (brown) for the standard model prediction including duality violation. If in future $\Delta \Gamma_s$ will be known with 
a precision of about $5\%$ both in theory and experiment, then the brown region will shrink to the orange one - here also the intrinsic
precision of the SM value will be reduced.
In other words, currently any measurement of $\Delta \Gamma_d$ outside the brown region will be a clear signal of new physics; in future
any measurement outside the orange region can be a signal of new physics.
The same logic is applied for the r.h.s. of  Fig.\ \ref{fig:duality_bounds}, where $a_{sl}^d$ and $a_{sl}^s$ are investigated simultaneously.
For $a_{sl}^s$ still any measurement outside the bounds in Eq.(\ref{asls3}) would be a clear indication of new physics. This bound is
in  Fig.\ \ref{fig:duality_bounds} denoted by the tiny brown region.
For  $a_{sl}^d$  the current experimental region is given by the green area, which is slightly smaller than the
orange region, which is indicating the theory prediction including duality violation. Future improvement in experiment and theory for 
$\Delta \Gamma_s$ will reduce the orange region to the dark blue one and then any measurement outside the dark blue region will be a clear signal of
new physics. 
\\
In addition we can ask if there are more observables that will be affected by the above discussed duality violations.
An obvious candidate is  the dimuon asymmetry, which depends on $a_{sl}^d$, $a_{sl}^s$ and $\Delta \Gamma_d$. This will be discussed
in Sec.~\ref{dimuon}. Another candidate is the the lifetime ratio $\tau (B_s^0) / \tau (B_d^0)$, where the dominant diagrams are very similar 
to the mixing ones, this observable will be studied further in  Sec.~\ref{lifetimeratio}.
\subsection{Duality bounds from the dimuon asymmetry}
\label{dimuon}%
The D0 collaboration has measured the like-sign dimuon asymmetry finding consistently deviations with the expected value from the Standard Model
\cite{Abazov:2010hv,Abazov:2010hj,Abazov:2011yk,Abazov:2013uma}.
The most recent experimental determination found a discrepancy of $3.0$ $\sigma$ when
interpreted as the result of CP violation in mixing and interference given in terms of the semileptonic asymmetries
$a^{s}_{sl}$, $a^d_{sl}$ and the life time difference $\Delta \Gamma_d$ respectively, as suggested by \cite{Borissov:2013wwa} and further
improved by \cite{niersteCKM}.
\\
Thus we want to investigate the possibility of explaining the discrepancy between theory and experiment as an
effect of duality violation.
The residual like-sign dimuon charge asymmetry $A_{CP}$ reads
\begin{eqnarray}
A_{CP}&=&C^s_{sl} a^{s}_{sl} + C^d_{sl} a^{d}_{sl} +  C_{int} \frac{\Delta \Gamma_d}{\Gamma_d} \, ,
\end{eqnarray}
with coefficients that can be determined using the information provided in \cite{Abazov:2013uma}, we also  include a further
correction factor in the interference contribution  $C_{int}$, as  suggested by \cite{niersteCKM}.
\begin{equation}
\begin{array}{|c||c|}
\hline
\mbox{Parameter}   & \mbox{Value}
\\
\hline
\hline
C^{d}_{sl}            &  0.220 \pm  0.018
\\
\hline
C^{s}_{sl}            & 0.157 \pm 0.013
\\
\hline
C_{int}             &  -0.040 \pm 0.013
\\
\hline
\end{array} 
\end{equation}
With this input we get a standard model estimate for $A_{CP}$ of
\begin{eqnarray}
A_{CP}^{\rm SM} &=&(-2.61 \pm 0.637)\cdot 10^{-4} \, .
\end{eqnarray}
Using our simple model for duality violation, see Eq. (\ref{model1}),  we get for the theory prediction
of  $A_{CP}$ after including duality violating effects
\begin{eqnarray}
A_{CP}=-2.61 (1-7.17 \delta)\cdot 10^{-4} \, .
\end{eqnarray}
This can be compared to the experimental result  provided by D0 \cite{Abazov:2013uma}
\begin{eqnarray}
A_{CP}=(-2.35 \pm 0.84 )\cdot 10^{-3} \, .
\end{eqnarray}
We find that there is an agreement between experiment and  theory if  $\delta$ lies
in the following region ($95$ $\%$ confidence level)
\begin{eqnarray}
-2.01  \leq \delta \leq -0.23 \, .
\end{eqnarray}
This is clearly out of the range found in Eq. (\ref{deltainterval}) from the direct constraints of mixing observables.
On the other hand we find with the allowed $\delta$-regions given  in Eq. (\ref{deltainterval}),
that $A_{CP}$ can be at most enhanced to
\begin{eqnarray}
 -4.52 \cdot 10^{-4} \leq A_{CP} \leq -1.06 \cdot 10^{-4} \, ,
\end{eqnarray}
which is considerably smaller than the experimental result.
This  excludes the possibility of explaining the current value for $A_{CP}$ as an effect of duality violation at the $2 \sigma$ level.

\subsection{Duality bounds from lifetime ratios} 
\label{lifetimeratio}%
\begin{figure}[htp]
\includegraphics[width=0.45\textwidth]{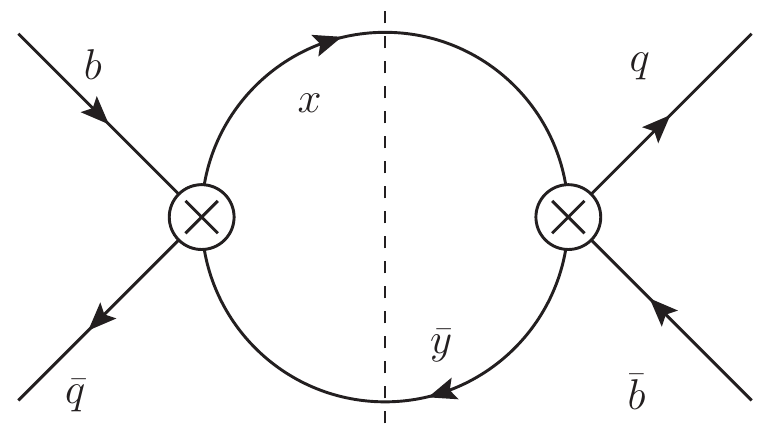}
\hfill
\includegraphics[width=0.45\textwidth]{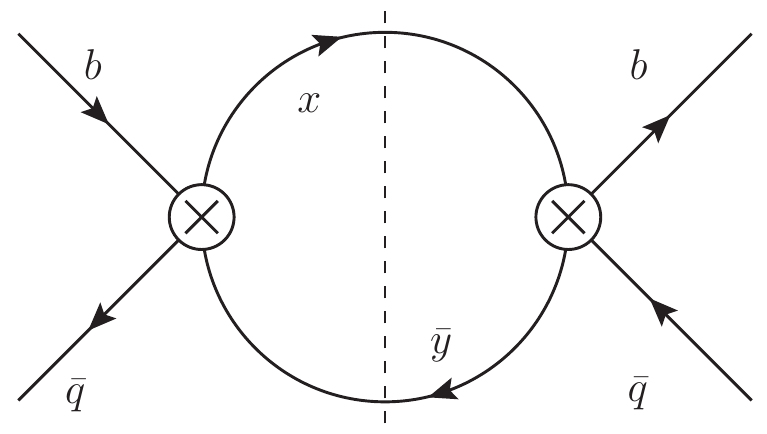}
\caption{Diagrams contributing to the  $\Gamma_{12}^q$ (l.h.s.) and diagrams contributing to the lifetime of a $B_q^0$-meson (r.h.s.).
         }
\label{fig:1-loop}
\end{figure}

Very similar diagrams to the ones in $\Gamma_{12}^q$ arise in the lifetime ratio $\tau (B_s^0) / \tau (B_d^0)$, see Fig.\ \ref{fig:1-loop}.
The obvious difference between the two diagrams is the trivial exchange of $b$ and $q$ lines at the right end of the diagrams. A more subtle and more 
important difference lies in the possible intermediate states, when cutting the diagrams in the middle. In the case of
lifetimes all possible intermediate states that can originate from a $x \bar{y}$ quark pair, can arise. In the case of mixing, we have only the
subset of all intermediate states into which both $B_q^0$ and $\bar{B}_q^0$ can decay. Thus one would expect that duality works better in the lifetimes
than in mixing.
Independent of this observation, our initial argument that the phase space for intermediate $c \bar{c}$-states is smaller than the one
for intermediate  $u \bar{c}$-states, which is again smaller that the  $u \bar{u}$-case, still holds.
Hence we assume that the $x \bar{y}$-loop for the lifetime ratio, has the same duality violating factor $\delta^{xy}$ as the
$x \bar{y}$-loop for $\Gamma_{12}^q$. 
It turns out that
the largest weak annihilation contribution to the $B_s^0$ lifetime is given by a $cc$-loop, while for the $B_d^0$ lifetime
a $uc$-loop is dominating. This observation tells us that duality will not drop out in the lifetime ratio, because
the dominating contributions for $B_s^0$ and $B_d^0$ are affected differently. 
Using our above model and modifying the $cc$-loop with a factor $1+4 \delta$ and the $uc$-loop with a factor $1+\delta$,
we get with the expressions in \cite{Lenz:2014jha,Uraltsev:1996ta,Neubert:1996we,Franco:2002fc}
\begin{equation}
\frac{\tau(B_s^0)}{\tau(B_d^0)} = 1.00050 \pm 0.00108 - 0.0225 \,\delta \, .
\label{lifetime}
\end{equation}
A detailed estimate of the theoretical  error is given in the Appendix \ref{app:B}. Unfortunately the standard model
prediction   relies strongly on lattice calculations that are already 15 years old \cite{Becirevic:2001fy}
and no update has been performed since then. For a more detailed discussion of the status of lifetime
predictions, see \cite{Lenz:2014jha}.
Nevertheless, one finds a big impact of the duality violating factor $\delta$ on the final result.
A value of $\delta = 1$ would have huge effects, compared to the central value within the standard model and its uncertainty.
\\
Our theory prediction can be compared to the  current experimental value for the lifetime ratio \cite{Amhis:2014hma}
\begin{equation}
\frac{\tau(B_s^0)}{\tau(B_d^0)} = 0.990 \pm 0.004 \, .
\label{lifetime_exp}
\end{equation}
If the tiny deviation between theory and experiment is attributed to duality violation, then 
we get an allowed range for $\delta $ of
\begin{eqnarray}
\delta & \in & [+0.24, +0.70] \, \, \mbox{(naive)} \, ,
\\
\delta & \in & [+0.13, +0.80 ] \, \, \mbox{(likelihood ratio $95\%$)} \, .
\label{boundslifetime}
\end{eqnarray}
There is currently a discrepancy of about \(2.5 \sigma\) between experiment (Eq.(\ref{lifetime_exp})) and theory 
(Eq.(\ref{lifetime})) and this
difference could stem from new physics or a sizeable duality violation of $\delta \approx 0.5$ in lifetimes, on the other hand we would 
like to remind the reader to the time evolution of the lifetime measurements shown in Fig.\ref{fig:lifetime_history}.
\begin{figure}[htp]
\includegraphics[width=0.9\textwidth]{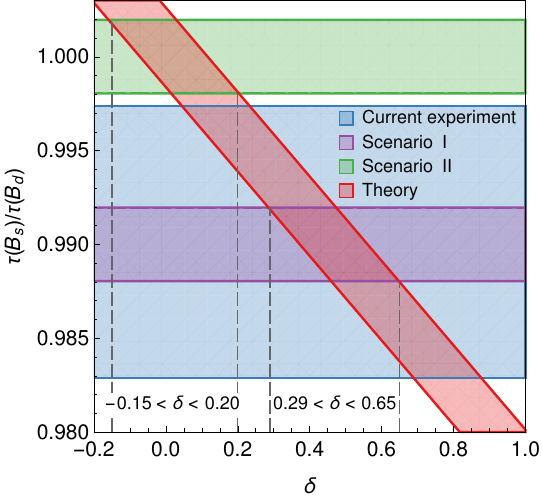}
\caption{Duality bounds extracted from the lifetime ratio $\tau (B_s^0) / \tau (B_d^0)$. The red band shows the theoretical
          expected value, see Eq.(\ref{lifetime}) of the lifetime ratio in dependence of the $\delta$. The current
          experimental bound is given by the blue region and the overlap of both gives the current allowed region
          $\delta$, indicated in Eq.(\ref{boundslifetime}). The future scenarios are indicated by the violet band (Scenario I)
          and the green band (Scenario II). Again the overlap of the future scenarios with the theory prediction
          gives future allowed regions for $\delta$ - in this figure the naive overlap of both regions is shown, this corresponds to 
          a linear addition of uncertainties and leads thus to slightly bigger ranges of $\delta$ compared to the text.
         }
\label{fig:lifetime_future_simple}
\end{figure}
The allowed region of the duality violating parameter $\delta$  can be read off Fig.\ \ref{fig:lifetime_future_simple},
where the current experimental bound from Eq.(\ref{lifetime_exp}) is given by the blue region and theory
prediction including hypothetical duality violation by the red region.
It goes without saying that 2.5 standard deviations is much too little to justify profound statements, thus we consider
next future scenarios where the experimental uncertainty of the lifetime ratio will be  reduced to $\pm 0.001$.
\begin{itemize}
\item Scenario I: the central value will stay at the current slight deviation from one:
      \begin{equation}
      \frac{\tau(B_s^0)}{\tau(B_d^0)}^{\rm Scenario \, I} = 0.990 \pm 0.001 \, .
       \end{equation}
       This scenario corresponds to a clear sign of duality violation or new physics in the lifetime ratio. 
       Assuming the first one, we get a range of $\delta$ of (see the violet region in   Fig.\ \ref{fig:lifetime_future_simple})
       \begin{eqnarray}
       \delta & \in & [0.38,0.56] \, \mbox{naive} \, ,
       \\
       \delta & \in & [0.34,0.60] \, \mbox{likelihood ratio 95\%} \, .
       \end{eqnarray}
       Thus the lifetime ratio requires large values of $\delta$. Our final conclusions depend now on the future
       developments of $\Delta \Gamma_s$. Currently  $\Delta \Gamma_s$ requires  small values of $\delta$, which is
       in contrast to scenario I. Thus we have to assume additional new physics effects - either in mixing or in lifetimes -
     that might solve the discrepancy.
       If in future the theory value of $\Delta \Gamma_s$ will go up sizeable or the experimental value will go down considerably, then
       mixing might also require a big value of $\delta$ and we then would have duality violation as a simple solution 
       for explaining discrepancies in both lifetimes and $B_s^0$ mixing.

\item Scenario II: the central value will go up to the standard model expectation:
      \begin{equation}
      \frac{\tau(B_s^0)}{\tau(B_d^0)}^{\rm Scenario \, II} = 1.000 \pm 0.001 \, ,
       \end{equation}
       In that case we will find only a  small allowed region for $\delta$ around zero, see the 
       green region in   Fig.\ \ref{fig:lifetime_future_simple}
       \begin{eqnarray}
       \delta & \in & [-0.07, 0.12] \, \mbox{naive} \, ,
       \\
       \delta  & \in & [-0.11, 0.15]
       \, \mbox{likelihood ratio 95\%} \, .
       \end{eqnarray}
       The above region is, however, still larger than the one  obtained from $\Delta \Gamma_s$. New lattice determinations
       of lifetime matrix elements might change this picture and in the end the lifetime ratio might also lead
       to slightly stronger duality violating bounds than $\Delta \Gamma_s$.
       Again our final conclusion depends on future developments related to $\Delta \Gamma_s$. If both experiment and theory
       for mixing stay at their current central values, we simply get very strong bounds on $\delta$. If theory or experiment 
       will change in future, when we could have indications for deviations in mixing, which have to be compared to the agreement of 
       experiment and theory for lifetimes in Scenario II.
\end{itemize}
In Section \ref{numericalupdate} we will discuss a possible future development of future theory predictions for mixing observables.
\\
Before we proceed let us make a comment about our duality model. In principle we also
could  generalise our duality ansatz, and modify the cc-loop with a factor \(1+\delta^{cc}\) 
and the uc-loop with a factor \(1+\delta^{uc}\), as we did in the mixing case. 
We get the following expression
\begin{equation}
\label{lifetimedeltas}
\frac{\tau(B_s^0)}{\tau(B_d^0)} = 1 + 0.0005 (1 - 13.4 \delta^{cc} + 8.92 \delta^{uc})
\end{equation}
Here one sees a pronounced cancellation of the $cc$ and the $uc$ contribution, if one allows $\delta^{cc}$ to be
of similar size as $\delta^{uc}$. This is, however, not what we expect from our phase space estimates for 
duality violation. Thus we use for the lifetime ratio only our model given in Eq.(\ref{lifetime}).

\section{Numerical Updates of Standard Model Predictions}
\label{numericalupdate}
\setcounter{equation}{0}

We have already pointed out that more precise values of $\Delta \Gamma_s$ are needed to derive 
more stringent bounds on duality violation in the $B$ system.
Very recently the  Fermilab MILC collaboration presented a comprehensive study
of the non-perturbative parameters that enter $B$-mixing \cite{Bazavov:2016nty}.\footnote{A first numerical 
analysis with this new inputs was already performed in 
       \cite{Blanke:2016bhf}; 
         but the authors put their emphasis on the implications for the correlation between $\Delta M_{s,d}$ and 
         $\varepsilon_K$ in models with constrained MFV and implications for $\Delta\Gamma_{s,d}$ have not been analysed.
         }
A brief summary of  their results reads:
\begin{itemize}
\item Improved numerical values for the non-perturbative matrix elements
      $\langle Q \rangle$, $\langle Q_S \rangle$, $\langle \tilde{Q}_S \rangle$,  
      $\langle R_0 \rangle$, $\langle R_1 \rangle$ and $\langle \tilde{R}_1 \rangle$ that are necessary for 
      $\Delta \Gamma_q$ and $\Delta M_q$. Hence we have numerical values for all operators that are arising up to dimension seven 
      in the HQE, up to $R_2$ and $R_3$, which are still unknown and can only be estimated by assuming vacuum insertion
      approximation.  
\item The results provide a very strong confirmation of vacuum insertion approximation.
      All their bag parameters turn out to be in the range of 0.8 to 1.2. 
      Sometimes in the literature different normalisations of the matrix elements are used, that lead to values of the 
      bag parameters which differ from one in vacuum insertion approximation, see e.g. the discussion in 
      \cite{Artuso:2015swg}. The definitions in \cite{Bazavov:2016nty} are all consistent with $B=1\pm 0.2$ in vacuum insertion
      approximation.
\item The  numerical values of $f_{B_q}^2 B$
      are larger than in most previous lattice calculations.
\end{itemize}
Based on these new results we perform a more aggressive  - compared to the recent study in \cite{Artuso:2015swg} - 
numerical analysis of the  SM predictions, where we try to  push the current theory uncertainties to the limits. 
In particular we will modify the predictions in \cite{Artuso:2015swg} by using
\begin{itemize}
\item Most recent values of the  CKM parameter from CKMfitter \cite{Charles:2004jd} (similar values can be 
      obtained from UTfit \cite{Ciuchini:2000de}).
\item New Fermilab MILC results for the bag parameters of $Q$, $\tilde{Q}_S$, $R_0$, $R_1$ and $\tilde{R}_1$. We do not try 
      to average with other lattice results, e.g. the values given by FLAG \cite{Aoki:2013ldr}.
\item Assume vacuum insertion approximation for $R_2$ and $R_3$ with a small uncertainty of $B = 1 \pm 0.2$. We note that 
      this is not clearly  justified yet and it has to be confirmed by independent determinations of the corresponding
       bag parameters.
\item Use results derived from equations of motion
       $\tilde{B}_{R_3} = 7/5 B_{R_3} - 2/5 B_{R_2}$
       and $\tilde{B}_{R_2} = - B_{R_2}$\cite{Beneke:1996gn}. 
\end{itemize}
All inputs are listed  in Appendix \ref{app:D}. We first note that the overall normalisation due to
$f_{B_q}^2 B$ seems to be considerably enhanced now, so we expect enhancements in $\Delta M_q$ and
$\Delta \Gamma_q$ that will cancel in the ratio. Moreover the uncertainty in the bag parameter ratio $\tilde{B}_S/B$ is larger than
e.g. in \cite{Artuso:2015swg}. On the other hand the dominant uncertainty due to $R_2$ and $R_3$ will now be dramatically  reduced.
\\
Putting everything together we get with the new parameters the following predictions for the two neutral $B$ systems, which are compared
with the more conservative theory predictions \cite{Artuso:2015swg} and the experimental values from HFAG  \cite{Amhis:2014hma}, that were
already given in Eq.(\ref{comparison1}).
\begin{equation}
\begin{array}{|c||r|r|r|}
\hline
\mbox{Observable} &  \mbox{SM - conservative}           & \mbox{SM - aggressive}                 & \mbox{Experiment}
\\
\hline
\hline
\Delta M_s        & ( 18.3 \pm 2.7)\,  \mbox{ps}^{-1 }  & ( 20.11 \pm 1.37)\,  \mbox{ps}^{-1 }&( 17.757 \pm 0.021) \, \mbox{ps}^{-1}
\\
\hline
\Delta \Gamma_s   & ( 0.088 \pm 0.020) \, \mbox{ps}^{-1 }&  ( 0.098 \pm 0.014) \, \mbox{ps}^{-1 }&( 0.082 \pm 0.006) \, \mbox{ps}^{-1}
\\
\hline
a_{sl}^s          & ( 2.22 \pm 0.27) \cdot 10^{-5}  & (2.27 \pm  0.25) \cdot 10^{-5}  & ( -7.5 \pm 4.1) \cdot 10^{-3}
\\
\hline
\frac{\Delta \Gamma_s }{\Delta M_s}
                 & 48.1 \, (1  \pm 0.173) \cdot 10^{-4}   & 48.8 \, ( 1 \pm 0.125) & 46.2 \, (1 \pm 0.073) \cdot 10^{-4}
\\
\hline
\hline
\Delta M_d        & ( 0.528 \pm 0.078)\,  \mbox{ps}^{-1 }  & ( 0.606 \pm 0.056) \,  \mbox{ps}^{-1 }&( 0.5055 \pm 0.0020) \, \mbox{ps}^{-1}
\\
\hline
\Delta \Gamma_d   & ( 2.61 \pm 0.59)\cdot 10^{-3} \, \mbox{ps}^{-1 }&  ( 2.99 \pm 0.52 )\cdot 10^{-3} \, \mbox{ps}^{-1 }& 0.66 (1 \pm 10) \cdot 10^{-3} \, \mbox{ps}^{-1}
\\
\hline
a_{sl}^d          & ( -4.7 \pm 0.6) \cdot 10^{-4}  & (-4.90 \pm 0.54) \cdot 10^{-4}    & ( -1.5 \pm 1.7) \cdot 10^{-3}
\\
\hline
\frac{\Delta \Gamma_d }{\Delta M_d}
                 & 49.4 \, (1  \pm 0.172) \cdot 10^{-4}   & 49.3 \, ( 1 \pm 0.149) & 13 \, (1 \pm 10) \cdot 10^{-3}
\\
\hline
\end{array}
\end{equation}
The new theory values for $\Delta M_q$ and $\Delta \Gamma_q$ are larger than the ones presented in \cite{Artuso:2015swg} and they are further
from experiment.  For the ratios $\Delta \Gamma_q / \Delta M_q$ and $a_{sl}^q$ the central values are only slightly 
enhanced. The overall error shrinks by about a factor of two for $\Delta M_s$ and also sizeably for $\Delta M_d$, $\Delta \Gamma_q$
and the ratios  $\Delta \Gamma_q / \Delta M_q$. For the semileptonic asymmetries the effect is less pronounced.
In Appendix \ref{app:D} a detailed analysis of the errors is given.
\\
It is now interesting to consider the ratios of the new SM predictions normalised to the experimental numbers.
\begin{eqnarray}
\frac{\Delta M_s^{\rm SM agr.}}{\Delta M_s^{\rm Exp}}           & = &   1.133   (1 \pm0.068) (1 \pm 0.0012) \,
\label{DMSMagr}
\\
                                                                & = &  1.133 (1 \pm 0.068) \, ,
\\
\frac{\Delta \Gamma_s^{\rm SM agr.}}{\Delta \Gamma_s^{\rm Exp}} & = &  1.20   (1 \pm 0.142) (1 \pm 0.073)
\label{DGSMagr}
\\
                                                                & = & 1.20 (1 \pm  0.16) \, ,
\\
\frac{\left( \frac{\Delta \Gamma_s }{\Delta M_s}\right)^{\rm Exp   .}}
     {\left( \frac{\Delta \Gamma_s }{\Delta M_s}\right)^{\rm SM agr.}}                       
                                                                & = &   0.947    (1 \pm 0.125) (1 \pm 0.073)
\label{DGDMSMagr}
\\
                                                                & = &  0.947 (1 \pm 0.145) \, ,
\\
\frac{\Delta M_d^{\rm SM agr.}}{\Delta M_d^{\rm Exp}}           & = & 1.201     (1 \pm  0.093) (1 \pm 0.0040)
\label{DMDMagr}
\\
                                                               & = & 1.20 (1 \pm 0.09) \, .
\end{eqnarray}
Here one clearly sees the enhancements of the mass differences, which are up to 20\% or more than two standard deviations above the 
experimental value. The decay rate difference $\Delta \Gamma_s$ is also enhanced by about 20\% above the measured value; due to larger 
uncertainties, this is statistically less significant. The dominant source for this enhancement is the new value of $\langle Q \rangle$.
The ratio $\Delta \Gamma_s / \Delta M_s$ is slightly lower than before, but still consistent with the corresponding experimental number.
\\
Taking the deviations above seriously, we can think about several possible interpretations:
\begin{enumerate}
\item Statistical fluctuations in the experimental results of the order of three standard deviations
      might explain the deviation in $\Delta \Gamma_s$, while the deviation in $\Delta M_s$ cannot be explained by 
      a fluctuation in the experiment.
\item Duality violations alone cannot explain these deviations, because they have no visible effects on $\Delta M_q$.
\item The lattice normalisation for $f_B^2 B$ is simply too high, future investigations will bring down the value
      and there is no NP in mixing. Currently there is no foundation for this possibility, but we try to leave no stone 
      unturned. Since $f_B^2 B$ cancels in the ratio of mass and decay rate difference, we can use the new values to
      give  the  most precise SM prediction of $\Delta \Gamma_s$  via
      \begin{eqnarray}
      \frac{\Delta \Gamma_s}{\Delta M_s} \cdot 17.757  \, \mbox{ps}^{-1}( \equiv \Delta M_s^{\rm exp}) 
      & = & 0.087 \pm 0.010 \, \mbox{ps}^{-1} \, .
      \end{eqnarray}
      Now the theory error is very close to the experimental one and it would be desirable to have more 
      precise values in theory and in experiment.
      In that case we also get an indication of the short-term perspectives for duality violating bounds. The above numbers
      indicate an  uncertainty of  $\pm 0.145$ for the ratio $\Delta \Gamma_s / \Delta M_s$, which corresponds - in the case of a perfect
      agreement of experiment and theory - to a bound on $\delta$ of $\pm 0.037$. This would already be a considerable improvement compared
      to the current situation. 
\item Finally the slight deviation might be a first hint for NP effects.  
      \begin{enumerate}
      \item To explain the deviation in the decay rate difference one needs new physics effects in tree 
            level decays, while deviation in $M_{12}$ might be solved by new physics effects in loop contributions.
      \item In principle one can think of the  possibility of new tree-level effects that modify both $\Delta \Gamma_s$ and $\Delta M_s$, 
            but which cancels in the ratio. $\Delta M_s$ is affected by a double insertion of the new tree-level operators.
            Following the strategy described in e.g.\cite{Bobeth:2014rda}, we found, however, that the possible effects on the mass
            difference are much too small. 
      \item Finally there is also the possibility of having a duality violation of about $20\%$  
            in $\Delta \Gamma_s$, while  the effect in $\Delta M_s$ is due to new physics in loops.
            This possibility can be tested in future by more precise investigations of the lifetime
            ratio $\tau (B_s^0) / \tau (B_d^0)$.
      \end{enumerate}
\end{enumerate} 
In order to draw any definite conclusions about these interesting possibilities, we need improvements in several sectors:
from experiment we need more precise values for $\Delta \Gamma_s$ and $\tau (B_s^0) / \tau (B_d^0)$. A first measurement of 
$\Delta \Gamma_d$ will also be very helpful. A measurement of the semileptonic asymmetries outside the duality-allowed regions
would already be a clear manifestation of new physics in the mixing system.
From the theory side we need (in ranked order)
\begin{enumerate}
\item A first principle determination of the dimension 7 operators $B_{R_{2,3}}$ and the corresponding colour-rearranged ones.
\item Independent non-perturbative determinations (lattice, sum rules) of the matrix elements of $Q$, $Q_S$, $\tilde{Q}_S$, $R_0$, $R_1$ and $\tilde{R}_1$.
      \item NNLO QCD calculations for the perturbative part of $\Gamma_{12}$.
\end{enumerate}
These improvements seem possible in the next few years and they might lead to a reduction of the theory error as low as $5\%$ and thus
might be
the path to a detection of new physics effects in meson mixing.

\section{D-mixing}
\label{Dmixing}
\setcounter{equation}{0}

D-mixing is by now experimentally well established and the values of the mixing parameters are 
quite well measured \cite{Amhis:2014hma}:
\begin{eqnarray}
x      & = &  (0.37 \pm 0.16) \cdot 10^{-2}  \, ,
\\
y      & = &  (0.66^{+0.07}_{-0.10}) \cdot 10^{-2}  \, .
\end{eqnarray}
Using $\tau (D^0) = 0.4101$ ps \cite{PDG}, this can be translated into
\begin{eqnarray}
\Delta M_D      =   \frac{x}{\tau (D^0)} & = & 0.00902 \, \mbox{ps}^{-1} \, ,
\\
\Delta \Gamma_D = 2 \frac{y}{\tau (D^0)} & = & 0.0322  \, \mbox{ps}^{-1} \, .
\end{eqnarray}
When trying to compare these numbers with theory predictions, we face the problem that it is not obvious 
if our theory tools  are also working in the $D$ system.
Till now the mixing quantities have been estimated via exclusive and inclusive approaches.
The exclusive approach is mostly based on phase space and $SU(3)_F$-symmetry arguments, 
see e.g. \cite{Falk:2001hx,Falk:2004wg}. Within this approach values for x and y of the order of  
1\% can be obtained. Thus, even if it is not a real first principle approach, this
method seems to be our best currently available tool to describe $D$ mixing.
Inclusive HQE calculations worked very well in the $B$ system, but their naive application to the 
$D$ system gives results that are  several orders of magnitude lower than the experimental result 
\cite{Georgi:1992as,Ohl:1992sr}. Hence it seems we are left with  some of the following options:
\begin{itemize}
\item The HQE is not valid in the charm system. This obvious solution might however, be challenged by the fact
      that the tiny theoretical $D$ mixing result is solely triggered by an extremely effective GIM cancellation
      \cite{Glashow:1970gm}, see e.g. the discussion in \cite{Bobrowski:2010xg}, and not by the 
      smallness of the first terms of the HQE expansion. A breakdown of the HQE in the charm system could best be 
      tested by investigating the lifetime ratio of $D$ mesons. From the theory side, the NLO QCD corrections
      have been determined for the lifetime ratio in \cite{Lenz:2013aua} and it seems that the experimental 
      measured values can be reproduced.
      To draw a definite conclusion about the agreement of experiment and theory for lifetimes and thus about the
      convergence of the HQE in the charm system, lattice evaluations of the
      unknown charm lifetime matrix elements are urgently needed. So this issue is currently unsettled.
\item Bigi and Uraltsev pointed out in 2000 \cite{Bigi:2000wn} that the extreme GIM cancellation in $D$ mixing 
      might be lifted by higher terms in HQE, i.e. the $1/m_c$-suppression of higher terms in the HQE is 
      overcompensated by a lifting of the GIM cancellation in higher order terms. There are indications for such an
      effect, see \cite{Bobrowski:2010xg,Bobrowski:2012jf}, but it is not clear whether the effect is large enough
      to explain the experimental mixing values. To make further progress in that direction we need the perturbative
      calculation of the dimension 9 and 12 terms of the OPE and an idea of how to estimate the matrix elements 
      of the arising D=9 and  D=12 operators. Hence this possibility is not ruled out yet.
\item The deviation of theory and experiment could of course also be due to new physics effects. Bounds on new
      physics models from determining their contributions to $D$ mixing, while more or less neglecting the standard
      model contributions  were studied e.g. in \cite{Golowich:2006gq}.
\end{itemize}
In this work we will investigate the related question, whether relatively small duality violating effects in inclusive 
charm decays could explain the deviation between experiment and the inclusive approach. We consider the decay 
rate difference $\Delta \Gamma_D$ for this task. According to the relation (see the derivation in \cite{Nierste:2009wg} or in Appendix \ref{app:C})
\begin{equation}
\Delta \Gamma_D \leq 2 |\Gamma_{12}|\, ,
\end{equation}
we will only study  $|\Gamma_{12}|$ and test whether it can be enhanced close to the experimental value of the
decay rate difference. This is of course only a necessary, but not a sufficient condition for an agreement
of experiment and theory. A complete answer would also require a calculation of $|M_{12}|$, which is beyond 
the scope of this work.
\\
$\Gamma_{12}$ consists again of three CKM contributions
\begin{equation}
\Gamma_{12} = - \left(    \lambda_s^2           \Gamma_{12}^{ss}
                       +2 \lambda_s \lambda_d \Gamma_{12}^{sd}
                       +  \lambda_d^2 \Gamma_{12}^{dd}
               \right) \; ,
\end{equation}
with the CKM elements $\lambda_d = V_{cd} V_{ud}^* $ and $\lambda_s =V_{cs} V_{us}^*  $.
Using again the unitarity of the CKM matrix ($\lambda_d + \lambda_s + \lambda_b = 0$)
we get
\begin{equation}
\Gamma_{12} = - \lambda_s^2  \left(  \Gamma_{12}^{ss}-2 \Gamma_{12}^{sd} + \Gamma_{12}^{dd} \right) 
              +2 \lambda_s \lambda_b \left( \Gamma_{12}^{sd} - \Gamma_{12}^{dd}\right)
              -  \lambda_b^2 \Gamma_{12}^{dd} \, .
\label{G12D}
\end{equation}
The CKM-factor have now a very pronounced hierarchy, they read
\begin{eqnarray}
 \lambda_s^2          & = &  4.82 \cdot 10^{-2} - 3.00 \cdot 10^{-6} I \, ,
\\
2 \lambda_s \lambda_b & = &  2.50 \cdot 10^{-5} + 5.91 \cdot 10^{-5} I \, ,
\\
\lambda_b^2           & = & -1.49 \cdot 10^{-8} + 1.53 \cdot 10^{-8} I \, .
\end{eqnarray}
The numerical values of the $\Gamma_{12}^{xy}$ can be expanded in powers of 
$\bar{z}_s = (\bar{m}_s(\bar{m}_c)/\bar{m}_c(\bar{m}_c))^2 \approx 0.0092$.
\begin{eqnarray}
\Gamma_{12}^{ss} & = & 1.8696 - 5.5231 \bar{z}_s - 13.8143 \bar{z}^2 + ... \bar{z}^3  + ... \, , 
\\
\Gamma_{12}^{sd} & = & 1.8696 - 2.7616 \bar{z}_s -  7.4906 \bar{z}^3 + ... \bar{z}^3    + ... \, ,
\\
\Gamma_{12}^{dd} & = & 1.8696 \, .
\end{eqnarray}
Looking at the expressions in Eq.(\ref{G12D}) we see an extreme GIM cancellation in the CKM-leading 
term, while the last term without any GIM cancellation is strongly CKM suppressed.
We get
\begin{eqnarray}
\Gamma_{12}^{ss}-2 \Gamma_{12}^{sd} + \Gamma_{12}^{dd} & = &  1.17 \bar{z}^2 - 59.5 \bar{z}^3 + ...   \, , 
\\
\Gamma_{12}^{sd} - \Gamma_{12}^{dd}  & = &  -2.76 \bar{z} + ... \, .
\label{G12Dpart}
\end{eqnarray}
Using our simplest duality violating model
\begin{eqnarray}
\Gamma_{12}^{ss} & \to & \Gamma_{12}^{ss} (1 + 4 \delta) \; ,
\label{adhoc1D}
\\
\Gamma_{12}^{sd} & \to & \Gamma_{12}^{sd} (1 +   \delta) \; ,
\label{adhoc2D}
\\
\Gamma_{12}^{dd} & \to & \Gamma_{12}^{dd} (1 + 0 \delta) \; ,
\label{adhoc3D}
\end{eqnarray}
we find
\begin{eqnarray}
\Gamma_{12}^{ss}-2 \Gamma_{12}^{sd} + \Gamma_{12}^{dd} & = &  1.17 \bar{z}^2 - 59.5 \bar{z}^3 + ... 
+ \delta \left(  3.7392  - 16.5692 z - 40.276 z^2 + ...\right)  \, , 
\\
\Gamma_{12}^{sd} - \Gamma_{12}^{dd}  & = &  -2.76 \bar{z} + ... 
+ \delta \left( 1.8696  - 2.7616 z - 7.4906 + ... \right) \, .
\label{G12Dpart2}
\end{eqnarray}
Eq.(\ref{G12Dpart2}) shows that our duality violating model completely lifts the GIM cancellation and that even tiny values 
of $\delta$ will lead to an overall result that is much bigger than the usual standard model predictions within the 
inclusive approach.
\begin{figure}[htp]
\centering
\includegraphics[width=0.9\textwidth]{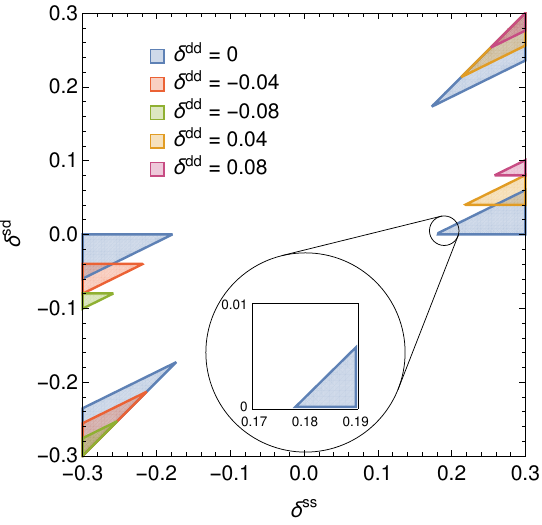}
\caption{95\% confidence limits on $\delta^{ss}$, $\delta^{sd}$ and $\delta^{dd}$ for the $D$ system from a comparison of the 
         experimentally 
         allowed region of $\Delta \Gamma_D$ with the theory prediction based on the HQE.
         The allowed regions for  the $\delta$s are shaded. Depending on the values of $\delta^{dd}$, different
         colours are used.  As expected for small values of $\delta$ the experimental value of 
         $\Delta \Gamma_D$ can not be reproduced. Thus the area in the centre is free.
         Starting from values of about $20\%$ on duality violation can explain the difference between
         experiment and HQE. To see more precisely, where the smallest possible value of $\delta$ lies,
         we have zoomed into the overlap region.
         }
\label{fig:D_general_limits}
\end{figure}
For our final conclusions we will use the generalised duality violating model
\begin{eqnarray}
\Gamma_{12}^{ss} & \to & \Gamma_{12}^{ss} (1 + \delta^{ss}) \; ,
\\
\Gamma_{12}^{sd} & \to & \Gamma_{12}^{sd} (1 + \delta^{sd}) \; ,
\\ 
\Gamma_{12}^{dd} & \to & \Gamma_{12}^{dd} (1 + \delta^{dd}) \; ,
\end{eqnarray}
with $\delta^{ss} \geq \delta^{sd} \geq \delta^{dd}$.
Next we test for what values of $\delta$ the inclusive approach can reproduce the experimental results for 
$\Delta \Gamma_D$. The corresponding allowed regions for $\delta^{ss,sd,dd}$ are given as shaded areas in Fig.
 \ref{fig:D_general_limits}. 
As expected, very small values of $\delta$ cannot give an agreement between HQE and experiment, surprisingly, however, 
values as low as  $\delta^{ss} \approx 0.18$ can explain the current difference.
So a duality violation of the order of $20\%$ in the HQE for the charm system
is sufficient to explain the huge discrepancy between a naive application of the HQE and the measured value for 
$\Delta \Gamma_D$.

\section{Summary and Conclusions}
\label{summary}
\setcounter{equation}{0}

In this paper we have explored the possibility of duality violations in heavy meson decays.
The study of such effects has a long tradition in flavour physics. Since the direct measurement of $\Delta \Gamma_s$ in 2012
by the LHCb collaboration huge duality violating effects are excluded \cite{Lenz:2012mb} by experiment.
But there is still space for duality violating effects of the order of $20\%$.
Because of the constantly improving experimental precision in flavour physics it is crucial to consider corrections of the order
of $20\%$ and to investigate whether, and how, such a bound can be improved.
\\
To do so, we introduced a simple parameterisation of duality violating effects, see Eq. (\ref{adhoc1}) - (\ref{adhoc3}),
 that relies solely on phase space arguments: the smaller the 
remaining phase space is in a heavy hadron decay, the larger duality violations might be.
In such a model, decay rate differences depend moderately on the duality violating parameter, $\delta$, whereas semi-leptonic asymmetries have a strong $\delta$ dependence, see Eq.(\ref{model1}). 
Currently we get the strongest bound on $\delta$ from Eq.(\ref{testHQE})
\begin{equation}
\frac{\left( \frac{\Delta \Gamma_s}{\Delta M_s} \right)^{\rm Exp}}{\left( \frac{\Delta \Gamma_s}{\Delta M_s}\right)^{\rm SM}} 
= 0.96 \pm 0.22 \, ,  \ \ |\delta| \lesssim 0.1.
\end{equation}
If the semileptonic asymmetries would agree with a similar precision between experiment and theory then the bound on $\delta$ would go down
to $\pm 0.01$. Unfortunately, the semileptonic asymmetries are not observed yet, and we have only experimental bounds. The same is true for 
the decay rate difference $\Delta \Gamma_d$.
Thus we use our bounds on $\delta$ from $\Delta \Gamma_s$ to determine the maximal possible size of $a_{sl}^q$ and $\Delta \Gamma_d$, 
if duality is violated. These regions are compared with current experimental ranges in Fig.\ref{fig:duality_bounds}.
Any measurement outside the region allowed by duality violation is a clear signal for new physics.
We also show a future scenario in which the duality violation is further constrained by more precise values of $\Delta \Gamma_s$ both in experiment and 
theory. 
\\
Duality violations would also affect the still unsolved problem of the dimuon asymmetry measured by the D0 collaboration, since it depends on 
$a_{sl}^d$, $a_{sl}^s$ and $\Delta \Gamma_d$. We found, however, that an agreement between experiment and theory for the dimuon asymmetry 
would require values of $\delta$ in the region of $-0.2$ to $-2.0$, which is considerably outside the allowed region found above. Taking only 
allowed values of $\delta$ we find that the theory prediction including duality violation is still an order of magnitude smaller than experiment.
Hence duality violation cannot explain the value of the dimuon asymmetry.  
\\
We have shown that the  duality violating parameter $\delta$ will also affect the lifetime ratio $\tau (B_s^0) / \tau (B_d^0)$, 
where we currently have a deviation of about 2.5 standard deviations between experiment and theory. Looking at the historical development 
of this ratio depicted in Fig. \ref{fig:lifetime_history} one might, however,  be tempted to assume a statistical fluctuation in the data.
Taking the current deviation seriously, it is either a hint for new physics or for a sizeable duality violations of the order of $ \delta \sim 0.5$,
which is inconsistent with our bounds on $\delta$ derived from $\Delta \Gamma_s$. Here a future reduction of the experimental error of  $\tau (B_s^0) / \tau (B_d^0)$
will give us valuable insight. We have studied two future scenarios in Fig. \ref{fig:lifetime_future_simple}, which would either point towards new physics
and duality violations or stronger bounds on duality violation. It is very important to note here that the theory prediction has a very strong dependence
on almost unknown lattice parameters. In particular, we can see from our error budget for the lifetime ratio in Appendix \ref{app:B} 
that any new calculation of the bag parameters \(\epsilon_{1,2}\) would bring large improvements in the theory prediction for $\tau (B_s^0)/ \tau(B_d^0)$.
\\ 
By now we already mentioned several times necessary improvements in both experiment and theory for mixing observables and in particular
for $\Delta \Gamma_s$.
Therefore we  presented an update of the SM predictions for the observables \(\Delta \Gamma, \Delta M,\) and \(a_{sl}\) in both the
 \(B_s^0\) and \(B_d^0\) systems, based on the recent Fermilab-MILC lattice results \cite{Bazavov:2016nty} for non-perturbative matrix elements, the latest 
CKM parameters from CKMfitter \cite{Charles:2004jd}, and an aggressive error estimate on the unknown bag parameters of dimension seven 
operators. With this input the current theory error in the mixing observables could be reduced by a factor of two for \(\Delta M_s\) or 1/3 for 
\(\Delta M_d, \Delta \Gamma_s,\) and \(\Delta M_s / \Delta \Gamma_s\). Thus we get for our  fundamental relation to establish the possible
size of duality violation
\begin{equation}
\frac{\left( \frac{\Delta \Gamma_s}{\Delta M_s} \right)^{\rm Exp}}{\left( \frac{\Delta \Gamma_s}{\Delta M_s}\right)^{\rm SM \, agr.}} 
= 0.95 \pm 0.15 \, .
\end{equation}
As expected, the overall uncertainty drops considerably, with a theory uncertainty almost compatible
with the experimental one  - thus demanding more precise experimental values of $\Delta \Gamma_s$.
On the other hand, we found in this new analysis that the central values of the mass differences and decay rate differences
are enhanced to values of about $20\%$ above the measurements with a significance of around 2 standard deviations.
To find out whether this enhancement is real, we need several ingredients: 1) an independent confirmation of the larger values 
of the matrix element $\langle Q \rangle$ found by \cite{Bazavov:2016nty}. 2) a first principle calculation of $ \langle R_{2,3} \rangle$ - triggered by the results
of  \cite{Bazavov:2016nty} we simply assumed small deviations from vacuum insertion approximation.  
If the new central values turn out to be correct, we will get profound implications for new physics effects
and duality violation in the $B$-system. For a further improvement
of the theory uncertainties NNLO-QCD corrections for mixing  have to be calculated.
\\
We finally focus on the charm system, where a naive application of  the HQE gives
results that are several orders of magnitude below the experimental values. We found the unexpected result
that duality violating effects as low as  $20\%$ could solve this discrepancy. 
Such a  result might have profound consequences on the applicability of the HQE. As a decisive test we suggest 
a lattice calculation of the matrix elements arising in the ratio of charm lifetimes. This ratio is free of any
GIM cancellation, which are very severe in mixing.

\section*{Acknowledgements}
This work is supported by the STFC grant of the IPPP.
T.J., M.K., and G.T.X. kindly acknowledge the financial support of Durham University, STFC, and CONACYT (Mexico) respectively.

\appendix

\section*{Appendices}
\renewcommand{\theequation}{\Alph{section}.\arabic{equation}}
\renewcommand{\thetable}{\Alph{table}}
\setcounter{section}{0}
\setcounter{table}{0}

\section{Error budget for lifetime ratio}
\label{app:B}
Varying each parameter individually, we get a total error of \(1.08 \cdot 10^{-3}\), 
which we note is smaller than the current experimental one by almost a factor of 4.
We present below the breakdown of contributions to the error for the lifetime ratio.
\begin{equation}
\begin{array}{|c||c|c|}
\hline
	& \tau(B^0_s) / \tau (B^0_d)\\
\hline
\hline
	\text{Central value} & 1.000503791\\
\hline
	\delta(\epsilon_1) & 0.00071\\
\hline
	\delta(\epsilon_2) & 0.00051\\
\hline
	\delta(f_{B_s}) & 0.00029\\
\hline
	\delta(\mu_G^2 (B_s)/\mu_G^2 (B_d)) & 0.00028\\
\hline
	\mu_\pi^2 (B_s) - \mu_\pi^2 (B_d)) & 0.00023\\
\hline
	\delta(f_{B_d}) & 0.00023\\
\hline
	\delta(c_3) & 0.00023\\
\hline
	\delta(\mu) & 0.00016\\
\hline
	\delta(B_1) & 0.00014\\
\hline
	\delta(\mu_G^2 (B_d)) & 0.00013\\
\hline
	\delta(B_2) & 0.00010\\
\hline
	\delta(c_G) & 0.000068\\
\hline
	\delta(m_b) & 0.000040\\
\hline
	\delta(|V_{cb}|) & 0.000025\\
\hline
	\delta(m_c) & 0.0000072\\
\hline
	\delta(\tau_{B_s}) & 0.0000013\\
\hline
	\delta(M_{B_s}) & 0.00000055\\
\hline
	\delta(M_{B_d}) & 0.00000025\\
\hline
	\delta(|V_{us}|) & 0.000000027\\
\hline
	\delta(\gamma) & 0.000000020\\
\hline
	\delta(|V_{ub} / V_{cb}|) & 0.000000010\\
\hline
\hline
	\sum \delta & 0.00108\\
\hline
\end{array}
\end{equation}

\newpage

\section{Proof of $\Delta \Gamma \leq 2 |\Gamma_{12}|$}
\label{app:C}
\setcounter{equation}{0}
In the $B$-system we get very simple expression for the mixing observables in terms of $M_{12}$ and $\Gamma_{12}$ because
one can make of the pronounced hierarchy $\Gamma_{12}/M_{12}$ and perform a Taylor expansion. In the $D$-system 
$\Delta \Gamma$ and $\Delta M$ are of the same order and one has to use the exact expression. One finds however,
$\Delta \Gamma \leq 2 |\Gamma_{12}|$, which gives us the opportunity to calculate only $\Gamma_{12}$ and to give an upper bound 
on $\Delta \Gamma$.
\\ 
We start with the two fundamental equations for the mixing observables:
\begin{eqnarray}
\label{eq:mixing1}
(\Delta M)^2 - \frac{1}{4} (\Delta \Gamma)^2 & = & 4 |M_{12}|^2 - |\Gamma_{12}|^2
\\
\label{eq:mixing2}
\Delta M \Delta \Gamma & = & 4 | M_{12}|  |\Gamma_{12}|  \cos \phi
\end{eqnarray}
where \(\phi = \arg(-M_{12} / \Gamma_{12})\).
Next we eliminate $\Delta M$ by substituting Eq.\ \ref{eq:mixing2} into Eq.\ \ref{eq:mixing1}, and then solve for 
$|M_{12}|$.
\begin{eqnarray}
\frac{16 |M_{12}|^2 |\Gamma_{12}|^2 \cos^2 \phi}{(\Delta \Gamma )^2} - \frac{1}{4} (\Delta \Gamma)^2 
&= & 
4 |M_{12}|^2 - |\Gamma_{12}|^2 
\\
|M_{12}|^2 \left( \frac{16 |\Gamma_{12}|^2 \cos^2 \phi}{(\Delta \Gamma)^2} - 4 \right) 
&= & 
\frac{1}{4} (\Delta \Gamma)^2 - |\Gamma_{12}|^2 
\\
|M_{12}|^2 &= & \frac{\frac{1}{4} (\Delta \Gamma)^2 - |\Gamma_{12}|^2}
                     {\left( \frac{16 |\Gamma_{12}|^2 \cos^2 \phi}{(\Delta \Gamma)^2} - 4 \right)} 
\label{eq:absMot}
\end{eqnarray}
Since $ M_{12} \geq 0$, we can say that the numerator and denominator on the r.h.s. of Eq. \ref{eq:absMot} 
must have the same sign.
\\
First, assume both terms are \(\geq 0\).
\begin{eqnarray}
\frac{1}{4} (\Delta \Gamma)^2 - | \Gamma_{12}|^2 \geq 0 
& & 
\frac{16 |\Gamma{12}|^2 \cos^2 \phi}{(\Delta \Gamma)^2} - 4 \geq 0 
\\
(\Delta \Gamma)^2 \geq 4 |\Gamma_{12}|^2 
& &  
(\Delta \Gamma)^2 \leq 4 \Gamma_{12}^2 \cos^2 \phi
\end{eqnarray}
These inequalities are only consistent in the case \(\cos^2 \phi = 1\) and \(\Delta \Gamma = 2 |\Gamma_{12}|\).
Now, assume both terms are \(\leq 0\).
\begin{eqnarray}
\frac{1}{4} (\Delta \Gamma)^2 - | \Gamma_{12}|^2 \leq 0 
& &
\frac{16 |\Gamma{12}|^2 \cos^2 \phi}{(\Delta \Gamma)^2} - 4 \leq 0 
\\
(\Delta \Gamma)^2 \leq 4 |\Gamma_{12}|^2 
& & 
(\Delta \Gamma)^2 \geq 4 \Gamma_{12}^2 \cos^2 \phi
\end{eqnarray}
Since \(0 \leq \cos^2 \phi \leq 1\), these inequalities are consistent for either
\begin{enumerate}
	\item \(\cos^2 \phi = 1\ \Rightarrow \Delta \Gamma = 2 |\Gamma_{12}|\)
	\item \(2 |\Gamma_{12}||\cos \phi| \leq \Delta \Gamma \leq 2 |\Gamma_{12}| \)
\end{enumerate}
We see that for either assumption, the inequality \(\Delta \Gamma \leq 2 |\Gamma_{12}|\) holds.

\newpage

\section{Numerical update with new lattice inputs}
\label{app:D}
\setcounter{equation}{0}

In this appendix we give details of the new numerical analysis done in Sec. \ref{numericalupdate}.
\subsection{Numerical Input}
For the inputs we use
\begin{equation}
\begin{array}{|c||c|c|}
\hline
\mbox{Parameter} & \mbox{This work} & \mbox{ABL 2015}
\\
\hline
\hline
f_{B_s} \sqrt{B} & 0.223 \pm 0.007 \, \mbox{GeV} & 0.215 \pm 0.015 \, \mbox{GeV}
\\
\hline
f_{B_d} \sqrt{B} &  0.185 \pm 0.008  \, \mbox{GeV}& 0175 \pm 0.012 \, \mbox{GeV}
\\
\hline
\tilde{B}_S/B (B_s^0)& 1.15 \pm 0.16 &  1.07 \pm 0.06
\\
\hline
\tilde{B}_S/B (B_d^0)& 1.17 \pm 0.24 &  1.04 \pm 0.12
\\
\hline
\tilde{B}_{R_0}/B (B_s^0)& 0.54 \pm 0.55 &  1.00 \pm 0.3
\\
\hline
\tilde{B}_{R_0}/B (B_d^0)& 0.35 \pm 0.80 &  1.00 \pm 0.3
\\
\hline
\tilde{B}_{R_1}/B (B_s^0)& 1.61 \pm 0.10 &  1.71 \pm 0.26
\\
\hline
\tilde{B}_{R_1}/B (B_d^0)& 1.72 \pm 0.15 &  1.71 \pm 0.26
\\
\hline
\tilde{B}_{\tilde{R}_1}/B (B_s^0)& 1.223 \pm 0.093 &  1.27 \pm 0.16
\\
\hline
\tilde{B}_{\tilde{R}_1}/B (B_d^0)& 1.31 \pm 0.14 &  1.27 \pm 0.16
\\
\hline
|V_{cb}|  &  0.04180^{+ 0.00033}_{-0.00068}  & 0.04117^{+ 0.00090 }_{-0.00114}
\\
\hline
|V_{ub}/V_{cb}|  &   0.0889 \pm 0.0019  & 0.0862 \pm 0.0044
\\
\hline
\gamma &  1.170^{+0.015}_{-0.035} & 1.171^{+0.017}_{-0.038} 
\\
\hline
|V_{us}| & 0.22542^{+0.00042}_ {-0.00031}   & 0.22548^{+0.00068}_{-0.00034}
\\
\hline
\end{array}
\end{equation}
\subsection{Central values}
With these  new input parameters we get the following predictions
\begin{equation}
\begin{array}{|c||c|}
\hline
\mbox{} & \mbox{This work}
\\
\hline
\hline
M_{12}^s       &  10.5 - 0.377 \cdot I
\\
\hline
M_{12}^d       &  0.214+0.215\cdot I
\\
\hline
\arg(M_{12}^s) &  -0.0375
\\
\hline
\arg(M_{12}^d) & 0.788
\\
\hline
\Gamma_{12}^s  &-0.0490+0.00207 \cdot I
\\
\hline
\Gamma_{12}^d  &-0.000950-0.00116 \cdot I
\\
\hline
\arg (\Gamma_{12}^s) & -0.0422
\\
\hline
\arg (\Gamma_{12}^d) & 0.886
\\
\hline
\Gamma_{12}^s / M_{12}^s    &  -0.00488 +0.0000227 \cdot I
\\
\hline
\Gamma_{12}^d / M_{12}^d    &  -0.00493-0.000490 \cdot I
\\
\hline
\end{array}
\end{equation}
from which we deduce the following observables
\begin{equation}
\begin{array}{|c||c|}
\hline
\mbox{Observable} & \mbox{This work}
\\
\hline
\hline
\Delta M_s     & 20.11 \pm  1.37 \,  \mbox{ps}^{-1}  
\\
\hline
\Delta M_d     & 0.606  \pm 0.056\, \mbox{ps}^{-1}    
\\
\hline
\Delta \Gamma_s &  0.098 \pm 0.014 \,  \mbox{ps}^{-1} 
\\
\hline
\Delta \Gamma_d &  0.00299 \pm 0.00052 \, \mbox{ps}^{-1} 
\\
\hline
\Re \left( \Gamma_{12}^s / M_{12}^s \right) & -0.00488  \pm 0.00061 
\\
\hline
\Re \left( \Gamma_{12}^d / M_{12}^d \right) & -0.00493  \pm 0.00073 
\\
\hline
\Im \left( \Gamma_{12}^s / M_{12}^s \right) & 0.0000227 \pm  2.50 \cdot 10^{-6}
\\
\hline
\Im \left( \Gamma_{12}^d / M_{12}^d \right) & -0.000490 \pm  0.000054 
\\
\hline
\pi - \arg \left( \Gamma_{12}^s / M_{12}^s \right) &  0.00466 \pm 0.00105
\\
& = ( 0.267   \pm 0.060)^{\circ}  
\\
\hline
\pi - \arg \left( \Gamma_{12}^d / M_{12}^d \right) & 0.0989  \pm 0.0233
\\
&  = (5.67   \pm 1.34)^{\circ}
\\
\hline
\end{array}
\end{equation}
\\
For completeness we give also
\begin{equation}
\begin{array}{|c||c|c|}
\hline
           & B_s^0              & B_d^0
\\
\hline
\hline
         c &   -48.65 \pm 6.10    &  -49.32   \pm 7.33
 \\
\hline
         a &  + 12.22 \pm 1.31    &  11.73    \pm 1.27
\\
\hline
         b &  + 0.77 \pm 0.10   &  0.23    \pm 0.04
\\
\hline
\end{array}
\end{equation}

\subsection{Error estimates}
We get now the following error estimates, compared to \cite{Artuso:2015swg}:
\\
\\
The mass difference $\Delta M_s$
\begin{equation}
\begin{array}{|c||c|c|}
\hline
\mbox{Parameter}          & \mbox{This work}      & \mbox {ABL2015}
\\
\hline
\hline
\delta (f_{B_s} \sqrt{B}) & 0.0635  & 0.139
\\
\hline
\delta (|V_{cb}|)           & 0.0240  & 0.049
\\
\hline
\delta (m_t)             & 0.0066   & 0.007
\\
\hline
\delta (\Lambda_{QCD})   & 0.0013   & 0.001
\\
\hline
\delta (\gamma)          & 0.0009   &  0.001
\\
\hline
\delta (m_b)             & 0.0005   & < 0.001
\\
\hline
\delta (|V_{ub}/V_{cb}|)   & 0.0004   & 0.001
\\
\hline
\hline
\sum \delta                   & 0.0682   & 0.148
\\
\hline
\end{array}
\end{equation}
\\
The mass difference $\Delta M_d$
\begin{equation}
\begin{array}{|c||c|c|}
\hline
\mbox{Parameter}          & \mbox{This work}      & \mbox {ABL2015}
\\
\hline
\hline
\delta (f_{B_d} \sqrt{B}) & 0.0872  &  0.137
\\
\hline
\delta (|V_{cb}|)         & 0.0240  & 0.049
\\
\hline
\delta (m_t)              & 0.0066   & 0.001
\\
\hline
\delta (\Lambda_{QCD})   & 0.0013   & 0.0
\\
\hline
\delta (\gamma)          & 0.0208   &  0.002
\\
\hline
\delta (m_b)             & 0.0005   & 0.0
\\
\hline
\delta (|V_{ub}/V_{cb}|)   & 0.0001   & 0.0
\\
\hline
\hline
\sum \delta                   & 0.0931   & 0.148
\\
\hline
\end{array}
\end{equation}
\\
The decay rate difference $\Delta \Gamma_s$
\begin{equation}
\begin{array}{|c||c|c|}
\hline
\mbox{Parameter}          & \mbox{This work}      & \mbox {ABL2015}
\\
\hline
\hline
\delta (\mu)              &  0.0889               & 0.084
\\
\hline
\delta (f_{B_s})          &  0.0635               & 0.139
\\
\hline
\delta (B_{R_2})          &  0.0604               & 0.148
\\
\hline
\delta (B_S)              & 0.0539                & 0.021
\\
\hline
\delta (B_{R_0})          & 0.0301                & 0.021
\\
\hline
\delta (|V_{cb}|)         & 0.0240                & 0.049
\\
\hline
\delta (\bar{z})          & 0.0109                & 0.011
\\
\hline
\delta (m_b)              & 0.0080                & 0.008
\\
\hline
\delta (B_{\tilde{R}_1})  & 0.0038                & 0.007
\\
\hline
\delta (m_s)              & 0.0024                & 0.001
\\
\hline
\delta (B_{R_3})          & 0.0023                & 0.002
\\
\hline
\delta (B_{R_1})          & 0.0018                & 0.005
\\
\hline
\delta (\gamma)           & 0.0010                & 0.001
\\
\hline
\delta (\Lambda_{QCD})    & 0.0010                & 0.001
\\
\hline
\delta (|V_{ub}/V_{cb}|)   & 0.0004                 & 0.001
\\
\hline
\delta (m_t)             & 0                      & <0.001
\\
\hline
\hline
\sum \delta                   & 0.1421                 & 0.228
\\
\hline
\end{array}
\end{equation}
\\
The decay rate difference $\Delta \Gamma_d$
\begin{equation}
\begin{array}{|c||c|c|}
\hline
\mbox{Parameter}          & \mbox{This work}      & \mbox {ABL2015}
\\
\hline
\hline
\delta (\mu)              & 0.0929                & 0.079
\\
\hline
\delta (f_{B_d})          & 0.0872               & 0.137
\\
\hline
\delta (\tilde B_S)       & 0.0809                & 0.04
\\
\hline
\delta (B_{R_2})          & 0.0623               & 0.144
\\
\hline
\delta (B_{R_0})          & 0.0533                & 0.025
\\
\hline
\delta (|V_{cb}|)         & 0.0240                & 0.049
\\
\hline
\delta (\gamma)           & 0.0233                & 0.002
\\
\hline
\delta (\bar{z})          & 0.0109                & 0.011
\\
\hline
\delta (m_b)              & 0.0076                & 0.008
\\
\hline
\delta (B_{R_3})          & 0.0023                & 0.005
\\
\hline
\delta (\Lambda_{QCD})    & 0.0009               & 0.001
\\
\hline
\delta (|V_{ub}/V_{cb}|)   & 0.0008                & 0.001
\\
\hline
\delta (B_{\tilde{R}_1})  & 0.0                                & 0.0
\\
\hline
\delta (m_d)              & --                                 & --
\\
\hline
\delta (B_{R_1})          & 0.0                                & 0.0
\\
\hline
\hline
\sum \delta               & 0.175                 & 0.227
\\
\hline
\end{array}
\end{equation}
\\
The real part of $\Gamma_{12}^s/M_{12}^s$
\begin{eqnarray}
\begin{array}{|c||c|}
\hline
\mbox{Parameter}             & \mbox{This work}\\
\hline
\hline
\delta (\mu)                 &  0.0889
\\
\hline
\delta (B_{R_2})             &  0.0604
\\
\hline
\delta (B_S)                 &  0.0539
\\
\hline
\delta (B_{R_0})             &  0.0301
\\
\hline
\delta (\bar{z})             &  0.0109
\\
\hline
\delta (m_b)                 &  0.0080
\\
\hline
\delta (m_t)                 &  0.0066
\\
\hline
\delta (\tilde{B}_{R_1})     &  0.0038
\\
\hline
\delta (m_s)                 &  0.0024
\\
\hline
\delta (\Lambda_{QCD})       &  0.0023
\\
\hline
\delta (B_{R_3})             &  0.0023
\\
\hline
\delta (B_{R_1})             &  0.0018 
\\
\hline
\delta (\gamma)              &  0.0001
\\
\hline
\delta (V_{ub}/V_{cb})       & 0 
\\
\hline
\delta (V_{cb})              &  0
\\
\hline
\hline
\sum \delta                  &   0.125
\\
\hline
\end{array}
\end{eqnarray}
\\
The real part of $\Gamma_{12}^d/M_{12}^d$
\begin{eqnarray}
\begin{array}{|c||c|}
\hline
\mbox{Parameter}             & \mbox{This work}\\
\hline
\hline
\delta (\mu)                &  0.0929
\\
\hline
\delta (B_S)                &  0.0809  
\\
\hline
\delta (B_{R_2})            &  0.0623
\\
\hline
\delta (B_{R_0})            &  0.0533
\\
\hline
\delta (\bar{z})            &  0.0109
\\
\hline
\delta (m_b)                &  0.0076
\\
\hline
\delta (m_t)                &  0.0066
\\
\hline
\delta (\gamma)             &  0.0025
\\
\hline
\delta (B_{R_3})            &  0.0023
\\
\hline
\delta (\Lambda_{QCD})      &  0.0022
\\
\hline
\delta (|V_{ub}/V_{cb}|)       & 0.0009
\\
\hline
\delta (\tilde{B}_{R_1})     &  0.0
\\
\hline
\delta (m_d)                 &  0.0
\\
\hline
\delta (B_{R_1})             &  0.0
\\
\hline
\delta (|V_{cb}|)            &  0.0
\\
\hline
\hline
\sum \delta                  &   0.149
\\
\hline
\end{array}
\end{eqnarray}
\\
The imaginary  part of $\Gamma_{12}^s/M_{12}^s$
\begin{eqnarray}
\begin{array}{|c||c|}
\hline
\mbox{Parameter}          & \mbox{This work}\\
\hline
\hline
\delta (\mu)              & 0.0946
\\
\hline
\delta (\bar{z})          & 0.0463 
\\
\hline
\delta (V_{ub}/V_{cb})    & 0.0211
\\
\hline
\delta (\gamma)           & 0.0118
\\
\hline
\delta (B_{R3})           & 0.0106
\\
\hline
\delta (m_b)              & 0.0101
\\
\hline
\delta (m_t)              & 0.0066
\\
\hline
\delta (B_{S})            & 0.0078
\\
\hline
\delta (\Lambda_{QCD})   & 0.0053
\\
\hline
\delta (B_{R2})          & 0.0039
\\
\hline
\delta (\tilde{B}_{R1})  & 0.0030
\\
\hline
\delta (B_{R0})          & 0.0026
\\
\hline
\delta (m_s)             & 0.0021
\\
\hline
\delta (B_{R1})          & 0.0002
\\
\hline
\delta (V_{cb})          & 0
\\
\hline
\hline
\sum \delta                   & 0.1098
\\
\hline
\end{array}
\end{eqnarray}
\\
The imaginary  part of $\Gamma_{12}^d/M_{12}^d$
\begin{eqnarray}
\begin{array}{|c||c|}
\hline
\mbox{Parameter}          & \mbox{This work}\\
\hline
\hline
\delta (\mu)              & 0.0937
\\
\hline
\delta (\bar{z})          & 0.0487
\\
\hline
\delta (|V_{ub}/V_{cb}|)    & 0.0215
\\
\hline
\delta (m_b)              & 0.0129
\\
\hline
\delta (B_{S})            & 0.0123
\\
\hline
\delta (B_{R3})           & 0.0115
\\
\hline
\delta (\gamma)           & 0.0105
\\
\hline
\delta (m_t)              & 0.0066
\\
\hline
\delta (\Lambda_{QCD})   & 0.0054
\\
\hline
\delta (B_{R0})          & 0.0049
\\
\hline
\delta (B_{R2})          & 0.0042
\\
\hline
\delta (\tilde{B}_{R1})  & 0.0
\\
\hline
\delta (m_d)             & 0.0
\\
\hline
\delta (B_{R1})          & 0.0
\\
\hline
\delta (|V_{cb}|)          &  0.0
\\
\hline
\hline
\sum \delta                   & 0.111
\\
\hline
\end{array}
\end{eqnarray}

\end{document}